\documentclass[pss]{wiley2sp} 
\usepackage{amsmath,amssymb}
\usepackage{verbatim}

\usepackage{graphicx,verbatim}
\usepackage{dcolumn}
\usepackage{bm}
\usepackage{sidecap}

\newcommand{\newwidth}{0.45\textwidth}
\newcommand{\newheight}{0.3\textwidth}
\newcommand{\newwidthprime}{0.15\textwidth}
\newcommand{\newheightprime}{0.2\textwidth}

\begin{document}

\title{Vibrationally coupled electron transport in single-molecule junctions: 
The importance  of electron-hole pair creation processes 
}

\titlerunning{Electron transport through molecular junctions: Role of pair creation processes}

\author{%
  R.\ H\"artle\textsuperscript{\Ast,\textsf{\bfseries 1}},
  U.\ Peskin\textsuperscript{\textsf{\bfseries 2}},
  M.\ Thoss\textsuperscript{\textsf{\bfseries 3}}}

\authorrunning{R.\ H\"artle, U.\ Peskin and M.\ Thoss}

\mail{e-mail
  \textsf{rh2613@columbia.edu}, Phone:
  +xx-xx-xxxxxxx, Fax: +xx-xx-xxx}

\institute{%
  \textsuperscript{1}\,Department of Physics, Columbia University, 538 West 120th Street, New York, NY 10027, USA.\\
  \textsuperscript{2}\,Schulich Faculty of Chemistry and the Lise Meitner Center for Computational Quantum Chemistry,\\
Technion-Israel Institute of Technology, Haifa 32000, Israel. \\
  \textsuperscript{3}\,Institut f\"ur Theoretische Physik und Interdisziplin\"ares Zentrum f\"ur Molekulare 
Materialien, 
Friedrich-Alexander-Universit\"at Erlangen-N\"urnberg,  
Staudtstr.\,7/B2, D-91058 Erlangen, Germany. }

\received{XXXX, revised XXXX, accepted XXXX} 
\published{XXXX} 


\abstract{%
\abstcol{%
Vibrationally coupled electron transport through single-molecule junctions is considered.
Reviewing our recent theoretical work, we show that electron-hole pair creation processes represent the key 
to understand the vibrational excitation characteristic of a single-molecule contact. 
Moreover, these processes can lead to a number of interesting
transport phenomena such as, for example, 
negative differential resistance, rectification, 
mode-se- }{ lective vibrational excitation and a pronounced temperature dependence of the electrical current. 
Thus, electron-hole pair creation processes 
are crucial to elucidate the basic mechanisms of vibrationally coupled electron transport through a single-molecule contact, 
despite the fact that these processes do not directly contribute to the electrical current that is 
flowing through the junction. 
  }}

\maketitle   

\section{Introduction}

The study of single-molecule junctions represents a vibrant and challenging 
field of research comprising fundamentally important aspects of quantum and many-body physics 
as well as the possibility for technological applications in nanoelectronic devices \cite{cuevasscheer2010}. 
Electron transport through a single-molecule junction can be understood in a similar way as 
through quantum dots, including electron-electron interactions \cite{Osorio2007} and/or quantum interference 
effects \cite{Guedon2011,Ballmann2012}. In addition, the vibrational degrees of freedom of a 
molecule play an important role \cite{Secker2010,Kim2011,Ballmann2012}, including, in particular,  
nonequilibrium effects such as current-induced vibrational excitation 
\cite{Galperin07,Pascual03,Schulze08,Hartle09,Saito2009,Huettel2009,Romano10,Hartle2010b,Natelson2011,Hartle2011b,Ballmann2012,Hartle2012}. 
This is in contrast to quantum dot systems \cite{Kiesslich2007}, where phonon degrees of freedom 
are often considered as being part of the environment rather than as active degrees of freedom, 
and is associated to the small size and mass of a molecular conductor.

The key to understand this nonequilibrium transport problem is to account for all relevant processes. 
Quite intuitively, this includes transport processes, where an electron is transferred from one electrode to another. 
In the presence of an external bias voltage, these processes give rise to an electrical current that is flowing through the molecule. 
However, these are not the only processes that occur in single-molecule junctions. Considering the simpler setup 
of a molecule adsorbed on a surface, it is well known that electron-hole pair creation processes, where an electron transfers 
from the substrate onto the molecule and back, constitute the most important relaxation mechanism for vibrational excitation \cite{Ueba2002}. 
Such pair creation processes do also occur in single-molecule junctions, 
although their importance for the respective nonequilibrium transport properties has been established only recently 
\cite{Schulze08,Hartle2010,Volkovich2011b,Hartle2011b,Hartle2011,Hartle2012,Ballmann2012}. 
In this article, we review the most important aspects of our work on this subject 
\cite{Hartle2010,Volkovich2011b,Hartle2011b,Hartle2011,HartlePhD,Hartle2012,Ballmann2012}. 
This includes a detailed understanding of the vibrational excitation 
characteristic, in particular the counter-intuitive phenomenon that larger levels of vibrational excitation are obtained 
for systems with weaker electronic-vibrational coupling \cite{Hartle2010b,Hartle2011}. In addition, pair creation processes 
may also have a substantial influence on the electrical transport properties of a single-molecule contact, 
giving rise, for example, to negative differential resistance, rectification \cite{Hartle2010b} and important implications 
for spectroscopical applications of single-molecule junctions \cite{HartlePhD}. 
Pair creation processes can even facilitate a mechanism to control 
the excitation level of specific vibrational modes selectively via the external bias voltage \cite{Hartle2010,Volkovich2011b} 
and, in the presence of destructive interference effects, of the electrical current via the temperature in the electrodes 
\cite{Hartle2011,Hartle2012}. Note that the latter has been experimentally verified only recently \cite{Ballmann2012}.

The article is organized as follows. In Sec.\ \ref{theorysec}, we introduce the model Hamiltonian and the nonequilibrium Green's function (NEGF)
approach \cite{Galperin06,Hartle,Hartle09,Volkovich2011b} that we use to describe electron transport through a single-molecule junction. 
Our results are presented in Sec.\ \ref{secResults}, where we set the stage by a discussion of the basic transport processes (Sec.\ \ref{SecBas})  
and electron-hole pair creation processes (Sec.\ \ref{SecElHole}), including 
a detailed analysis of the vibrational excitation characteristic of a generic molecular junction. 
Transport phenomena that are attributed to 
the influence of electron-hole pair creation processes are summarized in the last section, Sec.\ \ref{transphen}. 
Note that the presented data has already been published in Ref.\ \cite{HartlePhD}.

\section{Theory}
\label{theorysec}

\subsection{Model Hamiltonian}
\label{hamiltonian}

We consider a model for a molecular junction obtained by a partitioning of the overall system into
the molecule and the left and right leads.
The molecular part of a single-molecule junction (M) is often described 
by a set of discrete electronic states \cite{Galperin07,cuevasscheer2010}. 
The Hamiltonian of the corresponding electronic degrees of freedom can be written as \cite{Hartle2010b} 
\begin{eqnarray}
\label{hel} 
H_{\text{el}} &=& \sum_{m\in\text{M}} \epsilon_{m} c_{m}^{\dagger}c_{m}  \\ 
&& + \sum_{m<n\in\text{M}} U_{mn} (c_{m}^{\dagger}c_{m}-\delta_{m})
(c_{n}^{\dagger}c_{n}-\delta_{n}), \nonumber 
\end{eqnarray}
where $\epsilon_{m}$ denotes the energy of the $m$th 
molecular state and $c_{m}^{\dagger}$/$c_{m}$ the 
corresponding creation/annihilation operators. 
Thereby, the index $m$ distinguishes, in principle, 
different molecular orbitals, including the spin of the electrons. 
However, as the effects and mechanisms that are discussed in this review article 
do not explicitly involve spin degrees of freedom, we suppress them in the following. 
Coulomb interactions are described in the model Hamiltonian $H$ 
by Hubbard-like electron-electron interaction terms, 
$U_{mn}(c_{m}^{\dagger}c_{m}-\delta_{m}) (c_{n}^{\dagger}c_{n}-\delta_{n})$. 
Thereby, we account for the fact that the single-particle energies $\epsilon_{m}$ 
are determined with respect to a specific reference state of the molecule (\emph{e.g.}, 
the ground state of the uncharged molecule) \cite{Cederbaum74,Benesch06} and that, 
effectively, these energies include Coulomb interactions between the 
electrons of the reference system. Therefore, we distinguish  
occupied ($\delta_{m}=1$) and unoccupied states ($\delta_{m}=0$) of the reference state 
to avoid double counting of electron-electron interactions.

Besides the electronic degrees of freedom, nuclear motion and/or electronic-vibrational coupling plays 
a key role in transport through molecular conductors \cite{Galperin07,Hartle09,Osorio2010,Secker2010,Ballmann2012}, 
which is associated with their small size and mass. 
We take into account the vibrational degrees of freedom of a molecular conductor as harmonic oscillators that are 
linearly coupled to the electron (or hole) 
densities $(c_{m}^{\dagger}c_{m}-\delta_{m})$ \cite{Cederbaum74,Benesch06},   
\begin{eqnarray}
\label{Hvib}
H_{\text{vib}} &=& \sum_{\alpha} \Omega_{\alpha} a_{\alpha}^{\dagger}a_{\alpha} 
+ \sum_{m\alpha} \lambda_{m\alpha} Q_{\alpha} (c_{m}^{\dagger}c_{m}-\delta_{m}),
\end{eqnarray}
where the operator $a^{\dagger}_{\alpha}$ denotes the creation operator of the 
$\alpha$th oscillator with frequency $\Omega_{\alpha}$ and  
$Q_{\alpha}=a_{\alpha}+a_{\alpha}^{\dagger}$ the corresponding 
vibrational displacement operators. The respective 
coupling strengths are denoted by $\lambda_{m\alpha}$. 
Note that we identify the vibrational degrees of freedom of the junction as 
the normal modes of the aforementioned reference state. 
As a consequence, electronic-vibrational coupling is required to vanish for that state, 
which, similarly as for the electron-electron interaction terms, 
is ensured by the parameters $\delta_{m}$.

In most experiments on single-molecule junctions the molecule is contacted by metal electrodes. 
Accordingly, we describe the left (L) / right (R) electrode by 
a continuum of non-interacting electronic states  
\begin{eqnarray}
\label{hLR}
H_{\text{L/R}} &=& \sum_{k\in\text{L/R}} \epsilon_{k} c_{k}^{\dagger}c_{k}  
\end{eqnarray}
that are localized in the left/right lead. 
The coupling between the molecule and the leads is given by 
\begin{eqnarray}
\label{htun}
H_{\text{tun}} &=& \sum_{k\in\text{L,R};m\in\text{M}} ( V_{mk} c_{k}^{\dagger}c_{m} + \text{h.c.} ), 
\end{eqnarray} 
where the coupling matrix elements $V_{mk}$ 
determine the so-called level-width 
functions 
\begin{eqnarray}
\Gamma_{K,mn}(\epsilon)=2\pi\sum_{k\in K} 
V_{mk}^{*} V_{nk}\delta(\epsilon-\epsilon_{k})
\end{eqnarray}
with $K\in\{\text{L,r}\}$. 
Throughout this work, we assume semi-infinite tight-binding chains as models for the leads 
with an internal hopping parameter $\gamma=2$\,eV. 
The corresponding level-width functions are given by \cite{Cizek04}
\begin{eqnarray}
 \Gamma_{K,mn}(\epsilon) &=& \frac{ \nu_{K,m} \nu_{K,n}}{\gamma^{2}} \sqrt{4\gamma^{2}-(\epsilon-\mu_{K})^{2}},  
\end{eqnarray}
where, similar to $V_{mk}$, the parameters $\nu_{K,m}$ denote the coupling strength 
of state $m$ to lead $K$. 
In addition, 
we assume a symmetric drop of the bias voltage $\Phi$ at the contacts, 
\emph{i.e.}\ the chemical potentials in the left and the right lead 
are given by $\mu_{\text{L}}=e\Phi/2$ and $\mu_{\text{R}}=-e\Phi/2$, respectively. 
Furthermore, we set the Fermi energy of the leads to $\epsilon_{\text{F}}=0$\,eV. 
The Hamiltonian of the overall system is thus given by the sum 
\begin{eqnarray}
H &=& H_{\text{el}} + H_{\text{vib}} + H_{\text{L}} +H_{\text{R}} +H_{\text{tun}}. 
\end{eqnarray}

For the NEGF approach that we introduce in Sec.\ \ref{NEGFapproach}, 
it is expedient to remove the direct electronic-vibrational coupling terms in the Hamiltonian $H$ 
by the small polaron transformation \cite{Mitra04,Galperin06}:  
\begin{eqnarray}
\label{transformedHamiltonian}
\overline{H} &=& \text{e}^{S} H \text{e}^{-S} \,=\, \sum_{m} \overline{\epsilon}_{m} c_{m}^{\dagger}c_{m} 
+ \sum_{\alpha} \Omega_{\alpha} a^{\dagger}_{\alpha}a_{\alpha} \\
&& + \sum_{m<n} \overline{U}_{mn} (c_{m}^{\dagger}c_{m}-\delta_{m})
(c^{\dagger}_{n}c_{n}-\delta_{n}),   \nonumber\\
&& + \sum_{km}  (V_{mk} X_{m} 
c_{k}^{\dagger}c_{m} + \text{h.c.}) + \sum_{k} \epsilon_{k} c_{k}^{\dagger}c_{k},  \nonumber 
\end{eqnarray}
with 
\begin{eqnarray}
S &=& - i \sum_{m\alpha} (\lambda_{m\alpha}/\Omega_{\alpha})  ( c^{\dagger}_{m}c_{m} 
- \delta_{m}  )  P_{\alpha}, \\
X_{m} &=& \text{exp}[i\sum_{\alpha}(\lambda_{m\alpha}/\Omega_{\alpha}) P_{\alpha}], 
\end{eqnarray}
and $P_{\alpha}=-i(a_{\alpha}-a_{\alpha}^{\dagger})$. 
The effect of electronic-vibrational coupling is thus subsumed in 
a renormalization of the single-particle energies, 
$\overline{\epsilon}_{m}=\epsilon_{m}+(2\delta_{m}-1)\sum_{\alpha}(\lambda_{m\alpha}^{2}/\Omega_{\alpha})$, the 
electron-electron interaction strengths, $\overline{U}_{mn}=U_{mn}-2\sum_{\alpha}(\lambda_{m\alpha}\lambda_{n\alpha}/\Omega_{\alpha})$, 
and the molecule-lead coupling strengths that are dressed by the shift-operators $X_{m}$.

\subsection{Nonequilibrium Green's Function Approach}
\label{NEGFapproach}

All single-particle observables, such as, \emph{e.g.}, the population of levels 
or the electrical current that is flowing through the junction, can be calculated from the single-particle 
Green's functions of the system. For the electronic degrees of freedom of our present problem, 
they are given by 
 \begin{eqnarray}
G_{mn}(\tau,\tau') &=& -i \langle \text{T}_{c}
c_{m}(\tau)c_{n}^\dagger(\tau') \rangle_{H} \\ 
&=& -i \langle \text{T}_{c} c_{m}(\tau)X_m(\tau)c_{n}^\dagger(\tau') X^\dagger_{n}(\tau')\rangle_{\overline{H}}. \nonumber
\end{eqnarray}
where the indices $H/\overline{H}$ indicate the Hamiltonian that 
is used to evaluate the respective expectation values. 
We employ the ansatz \cite{Galperin06,Hartle,Hartle09,Volkovich2011b}: 
\begin{eqnarray}
\label{decoupling}
G_{mn}(\tau,\tau') &\approx& \bar{G}_{mn}(\tau,\tau') \langle \text{T}_{c} X_m(\tau)X_{n}^\dagger(\tau') \rangle_{\overline{H}}, 
\end{eqnarray}
to calculate 
these Green's functions with 
$\bar{G}_{mn}(\tau,\tau')= -i \langle \text{T}_{c} c_{m}(\tau)c_{n}^\dagger(\tau') \rangle_{\overline{H}}$ 
and $\text{T}_{c}$ the time-ordering operator on the Keldysh contour.  
The ansatz (\ref{decoupling}) represents an effective factorization of 
the single-particle Green's functions $G_{mn}$ into 
a product of the electronic Green's functions, $\bar{G}_{mn}$, 
and a correlation function of shift operators, 
$\langle \text{T}_{c} X_m(\tau)X_{n}^\dagger(\tau') \rangle_{\overline{H}}$. 
This is justified, if the time scales for electronic processes and vibrational motion in the junction are different, 
and is conceptually similar to the Born-Oppenheimer approximation \cite{Domcke04}.

We use an equation of motion technique to determine the single-particle Green's functions of the present problem 
\cite{Galperin06,Hartle,Hartle09,Volkovich2011b}. 
The equation of motion for the electronic part of the Green's function reads 
\begin{eqnarray}
\label{eleom}
 (i\partial_{\tau}-\overline{\epsilon}_{m}) \bar{G}_{mn}(\tau,\tau') (-i\partial_{\tau'}-\overline{\epsilon}_{n}) &=& \\
 &&\hspace{-4cm} \delta(\tau,\tau') \delta_{mn} (-i\partial_{\tau'}-\overline{\epsilon}_{n}) \nonumber\\
 &&\hspace{-4cm} + \Sigma_{\text{Coul},mn}(\tau,\tau') + \sum_{K\in\{\text{L,R}\}} \Sigma_{K,mn}(\tau,\tau'),  \nonumber
\end{eqnarray}
where the self-energy contributions due to the coupling 
of the molecule to the left and the right leads are given by 
\begin{eqnarray}
\Sigma_{\text{L/R},mn}(\tau,\tau')&=&\\
&&\hspace{-1cm}\sum_{k\in \text{L/R}} 
V_{mk}^{*} V_{nk} g_{k}(\tau,\tau')\langle \text{T}_{c} X_{n}(\tau')X_{m}^\dagger(\tau) \rangle_{\overline{H}} \nonumber
\end{eqnarray} 
with $g_{k}(\tau,\tau')$ the free Green's function associated with lead state $k$. 
The effect of electron-electron interactions is subsumed in the self-energy 
$\Sigma_{\text{Coul},mn}(\tau,\tau')$. We describe the latter in terms of the 
elastic co-tunneling approximation \cite{Groshev,Stafford09,Hartle09,Volkovich2011b}, 
where $\Sigma_{\text{Coul},mn}(\tau,\tau')$ is replaced by the self-energy 
$\Sigma^{0}_{\text{Coul},mn}(\tau,\tau')$ describing electron-electron interactions 
in the isolated molecule (\emph{i.e.}\ where $V_{mk}=0$), facilitating 
a nonperturbative description of electron-electron interactions.

The correlation functions of the shift operators, $\langle \text{T}_{c} X_m(\tau)X_{n}^\dagger(\tau')
\rangle_{\overline{H}}$, which represent the second part of the factorized Green's function $G_{mn}$ 
(cf.\ Eq.\ (\ref{decoupling})), can be determined employing  
a second-order cumulant expansion in the dimensionless coupling parameters 
$\frac{\lambda_{m\alpha}}{\Omega_{\alpha}}$ \cite{Hartle09,Volkovich2011b}
\begin{eqnarray}
&&\langle \text{T}_{c} X_m(\tau)X_{n}^\dagger(\tau')
\rangle_{\overline{H}}= \\
&&\text{e}^{\sum_{\alpha\alpha'}i\frac{\lambda_{m\alpha}\lambda_{n\alpha'}}{\Omega_{\alpha}
\Omega_{\alpha'}}D_{\alpha\alpha'}(\tau,\tau') -i\frac{\lambda_{m\alpha}\lambda_{m\alpha'}+\lambda_{n\alpha}\lambda_{n\alpha'}}{2\Omega_{\alpha}
\Omega_{\alpha'}}D_{\alpha\alpha'}(\tau,\tau) }, \nonumber
\end{eqnarray}
with the momentum correlation functions 
\begin{eqnarray}
D_{\alpha\alpha'}=-i\langle \text{T}_{c} P_{\alpha}(\tau)P_{\alpha'}(\tau')\rangle_{\overline{H}}.  
\end{eqnarray}
These single-particle Green's functions can also be determined evaluating their equations of motion 
\begin{eqnarray}
\label{vibeom}
&&\frac{1}{4\Omega_{\alpha}\Omega_{\alpha'}} (-\partial^{2}_{\tau}-\Omega_{\alpha}^{2}) D_{\alpha\alpha'}(\tau,\tau') 
(-\partial^{2}_{\tau'}-\Omega_{\alpha'}^{2}) = \\ 
&&\delta(\tau,\tau') \delta_{\alpha\alpha'} 
(-\partial^{2}_{\tau'}-\Omega_{\alpha'}^{2}) \frac{1}{2\Omega_{\alpha'}} + \Pi_{\text{el},\alpha\alpha'}(\tau,\tau'). \nonumber
\end{eqnarray}
The corresponding self-energy matrix $\Pi_{\text{el},\alpha\alpha'}$ 
is evaluated up to second order 
in the molecule-lead coupling \cite{Hartle09,Volkovich2011b} 
\begin{eqnarray}
\label{Piel}
&&\Pi_{\text{el},\alpha\alpha'}(\tau,\tau')=-i\sum_{mn}\frac{\lambda_{m\alpha}\lambda_{n\alpha'}}{\Omega_{\alpha} \Omega_{\alpha'}} \times \\
&&(\Sigma_{mn}(\tau,\tau')\bar{G}_{nm}(\tau',\tau)+\Sigma_{nm}(\tau',\tau)\bar{G}_{mn}(\tau,\tau')). \nonumber
\end{eqnarray}
Since $\Pi_{\text{el},\alpha\alpha'}$ depends on the electronic self-energies 
$\Sigma_{mn}=\Sigma_{\text{L},mn}+\Sigma_{\text{R},mn}$ and Green's functions $\bar{G}_{mn}$, 
the evaluation of Eqs.\ (\ref{eleom}) and (\ref{vibeom}) requires 
an iterative self-consistent solution scheme \cite{Hartle09,Volkovich2011b}.

\subsection{Observables of interest}
\label{currentandvibex}

To characterize electron transport through a single-molecule junction, we analyze two different 
observables of interest as functions of the applied bias voltage $\Phi$: 
the average vibrational excitation and the electrical current flowing through the junction. 
The average level of excitation of vibrational mode $\alpha$ is given by 
\begin{eqnarray}
\label{formulavibex}
\langle a^\dagger_{\alpha} a_{\alpha} \rangle_{H} &=& \langle a^\dagger_{\alpha} a_{\alpha} \rangle_{\overline{H}}  \\
&&\hspace{-0.25cm} + \sum_{mn}\frac{\lambda_{m\alpha}\lambda_{n\alpha}}{\Omega_{\alpha}^2} \langle (c_{m}^{\dagger}c_{m}-\delta_{m}) 
(c_{n}^{\dagger}c_{n}-\delta_{n}) \rangle_{\overline{H}}.  \nonumber
\end{eqnarray} 
where the latter terms represent the contribution from polaron-formation, which  
is associated with charging of the molecular bridge. 
It can be computed from 
the vibrational Green's function $D_{\alpha\alpha}$ according to \cite{Volkovich2011b}
\begin{eqnarray}
\label{formulavibex2}
\langle a^\dagger_{\alpha} a_{\alpha} \rangle_{H}&\approx&-\frac{1}{2}\text{Im}\left[D^{<}_{\alpha\alpha}(t=0)\right] -\frac{1}{2} \\ 
&&\hspace{-1.5cm} +\sum_{m}\frac{\lambda_{m\alpha}^{2}}{\Omega_{\alpha}^2} \left(\text{Im}[\bar{G}^<_{mm}(t=0)] - \delta_{m} \right) \nonumber \\
&&\hspace{-1.5cm} -2\sum_{m<n}\frac{\lambda_{m\alpha}\lambda_{n\alpha}}{\Omega_{\alpha}^2} \left( 
\text{Im}[\bar{G}^<_{mn}(t=0)]\text{Im}[\bar{G}^<_{nm}(t=0)]  \right. \nonumber\\
&&\hspace{-1.5cm} - \left. \text{Im}[\bar{G}^<_{mm}(t=0)] - \delta_{m} \right) 
\left(\text{Im}[\bar{G}^<_{nn}(t=0)] - \delta_{n} \right), \nonumber 
\end{eqnarray}
where the electronic Green's function $\overline{G}_{mn}$ and the Hartree-Fock-like factorization 
\begin{eqnarray}
 \langle c_{m}^{\dagger}c_{m} c_{n}^{\dagger}c_{n} \rangle_{\overline{H}} &\approx&   
\langle c_{m}^{\dagger}c_{m} \rangle\langle c_{n}^{\dagger}c_{n} \rangle-\langle c_{m}^{\dagger}c_{n} \rangle\langle c_{n}^{\dagger}c_{m} \rangle 
\end{eqnarray}
for $m\neq n$ 
is used to represent the contribution from polaron-formation.

The current through lead $K$, $I_{K}$, is determined by the number of electrons entering or leaving the  
lead in a given time interval ($K\in\lbrace \text{L,R}\rbrace$)
\begin{eqnarray}
\label{firstcurrent}
I_{K} &=& -2e \frac{\text{d}}{\text{d} t} 
\sum_{k\in K} \langle c^{\dagger}_{k} c_{k} \rangle_{\overline{H}} . 
\end{eqnarray}
Here, the constant ($-e$) denotes the electron charge and the factor $2$ accounts 
for spin-degeneracy. 
With the self-energies $\Sigma_{\text{L/R},mn}$ and 
Green's functions $G_{mn}$ (cf.\ Sec.\ \ref{NEGFapproach}), 
the current can be calculated employing the Meir-Wingreen-like formula \cite{Meir92,Hartle,Hartle09}
\begin{eqnarray}
\label{currentformula}
&& I_{K} = \\
&& 2e\int\frac{\text{d}\epsilon}{2\pi}\, \sum_{mn} 
\left( \Sigma_{\text{K},mn}^{<}(\epsilon)\bar{G}^{>}_{nm}(\epsilon)-\Sigma_{\text{K},mn}^{>}(\epsilon)\bar{G}^{<}_{nm}(\epsilon) \right) . \nonumber
\end{eqnarray}
Note that the NEGF approach is current conserving, \emph{i.e.}\ $I_{\text{L}}=-I_{\text{R}}=I$.

\subsection{Vibrations in thermal equilibrium}
\label{thermalequSection}

In addition to the nonequilibrium treatment discussed above,
we also employ a simpler 'equilibrium' description, where we neglect the 
vibrational self-energy, \emph{i.e.}\ $\Pi_{\text{el},\alpha\alpha'}(t,t')=0$. 
This is a commonly used approximation \cite{Mitra04,Galperin06,Hartle,Hartle2010b} and 
constitutes a powerful tool to elucidate and analyze nonequilibrium effects. 
Within this approximation, 
the vibrational degrees of the junction are effectively 
confined to the state they would acquire in thermal equilibrium 
at an effective temperature $k_{\text{B}}T$. This means that, in contrast to a 
full nonequilibrium calculation, tunneling electrons may, indeed, excite or 
deexcite the vibrational degrees of freedom of the junction but that such changes of the vibrational state 
decay instantly such that an electron that is tunneling subsequently through the junction 
finds the vibrational mode in the same equilibrium state as before.

\section{Results}
\label{secResults}

We study the importance of resonant electron-hole pair creation processes for vibrationally coupled electron 
transport in single-molecule junctions in three steps of increasing complexity. 
First, in Sec.\ \ref{SecBas}, we introduce the basic transport processes and mechanisms that occur in 
the resonant transport regime of a molecular junction. To this end, we discuss the current-voltage characteristics 
of two generic model systems of molecular junctions that include a single and two electronic states. 
This provides the basis for our discussion in Sec.\ \ref{SecElHole}, where we introduce 
electron-hole pair creation processes on the very same grounds as transport processes. 
Thereby, we analyze the vibrational excitation characteristics of the model systems that we already discussed 
in Sec.\ \ref{SecBas}. It is shown that electron-hole pair creation processes are crucial to understand 
the (current-induced) levels of vibrational excitation in a single-molecule junction, including 
the respective electrical transport properties. This is exemplified in more detail 
in Sec.\ \ref{transphen}, where we discuss various transport phenomena that can be traced back 
to the influence of electron-hole pair creation processes. 
Note that the validity of the results, which are presented in the following and 
based on the NEGF scheme outlined in Sec.\ \ref{NEGFapproach}, have been corroborated 
by comparison to Born-Markov master equation approaches \cite{Hartle09,Volkovich2011b,HartlePhD} and a 
numerically exact wave-propagation scheme \cite{Thoss2011}.

\begin{table}
\begin{center}
\begin{tabular}{|*{7}{c|}}
\hline \hline
model & $\epsilon_{1}$ &  $\nu_{\text{L},1}$ & $\nu_{\text{R},1}$ & $\gamma$ & $\Omega_{1}$ 
&  $\lambda_{11}$   \\ 
\hline 
E1V1 & 0.6 & 0.1 & 0.1 & 3 & 0.1 & 0.06 \\
BAND & 0.15 & 0.02 & 0.02 & 0.2 & 0.1 & 0.06 \\
REC & 0.6 & 0.1 & 0.03 & 3 & 0.1 & 0.06  \\
RECBD & 0.15 & 0.02 & 0.006 & 0.2 & 0.1 & 0.06 \\
\hline \hline
\end{tabular}
\end{center}
\caption{\label{tableI}
Model parameters for molecular junctions with    
a single electronic state (energy values are given in $\mathrm{eV}$). 
The temperature in the leads is assumed to be $T=10$\,$\mathrm{K}$ 
throughout the article. Note that 
the parameters of all the model systems that we discuss in this review article 
fulfill the antiadiabatic condition $\Gamma_{\text{L/R},11}<\Omega_{1}$ and that they 
are in line with previous experimental 
\cite{Boehler04,Elbing05,Sapmaz05,Tao2006,Natelson2006,Osorio2010,Ballmann2010,Secker2010,Ballmann2012} and 
theoretical findings \cite{Xue01,Evers04,Benesch06,Benesch08,Benesch2009,Frederiksen2010}. 
} 
\end{table}

\begin{table*}
\begin{center}
\begin{tabular}{|*{15}{c|}}
\hline \hline
model & $\epsilon_{1}$ &  $\epsilon_{2}$&  $U_{12}$ & $\nu_{\text{L},1}$ & $\nu_{\text{L},2}$ & $\nu_{\text{R},1}$ & $\nu_{\text{R},2}$ 
& $\gamma$ & $\Omega_{1}$& $\Omega_{2}$ & $\lambda_{11}$ & $\lambda_{21}$ & $\lambda_{12}$ & $\lambda_{22}$ \\ \hline 
E2V1 & 0.15 & 0.8 & 0 & 0.1 & 0.1 & 0.1  & 0.1 & 3 & 0.1 & -- & 0.06 & -0.06 & -- & -- \\
SPEC & 0.15 & 0.8 & 0 & 0.1 & 0.1 & 0.03  & 0.03 & 3 & 0.1& --  & 0.06 & -0.06 & -- & -- \\
MSVE & 0.65 & -0.575 & 0 & 0.1 & 0.1 & 0.03  & 0.03 & 2 & 0.15 & 0.2  & 0.09 & 0 & 0 & 0.12 \\
INT & 0.5123 & 0.5126  & 0.0248 & 0.01 & 0.01 & 0.01  & -0.01 & 0.25 & 0.005 & --  & 0.0248 & 0.025 & -- & -- \\
\hline \hline
\end{tabular}
\end{center}
\caption{\label{tableII}
Model parameters for molecular junctions with  
two electronic states (energy values are given in $\mathrm{eV}$). 
}
\end{table*}

\subsection{Basic transport processes} 
\label{SecBas}

\begin{figure} 
\begin{center}
\begin{tabular}{l}
\resizebox{\newwidth}{\newheight}{
\includegraphics{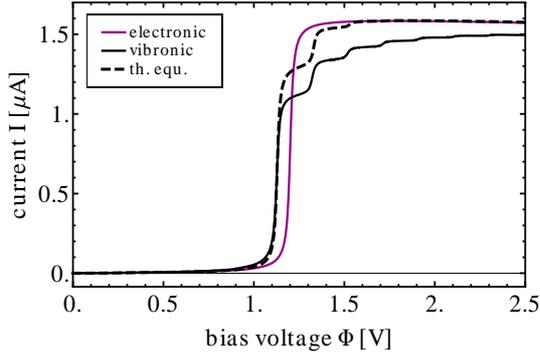}
}
\\
\end{tabular}
\end{center}
  \caption{\label{K21-current} Current-voltage characteristics of junction E1V1 (cf.\ Tab.\ \ref{tableI} for the corresponding model parameters). 
 }
\end{figure}

To discuss the basic transport processes that occur in single-molecule junctions, 
we consider first a simple model for a molecular junction 
with a single electronic state that is coupled 
to a single vibrational mode (model E1V1, see Tab.\ \ref{tableI} 
for a detailed list of all model parameters). 
The corresponding current-voltage characteristic 
is shown in Fig.\ \ref{K21-current}, where we distinguish three different scenarios: 
$i$) the electronic scenario (solid purple line), where we neglect the effect of electronic-vibrational coupling, 
$ii$) the vibronic scenario (solid black line), where we fully account for the coupling between the electronic state and the vibration,   
and $iii$) the thermally equilibrated scenario (dashed black line), where electronic-vibrational is accounted for but the vibration 
is restricted to the state it acquires in thermal equilibrium at $T=10$\,K (which is effectively its ground state).

\begin{figure*}
\begin{center}
\begin{tabular}{llll}
(a)&(b)&(c)&(d)\\
\resizebox{\newwidthprime}{\newheightprime}{
\includegraphics{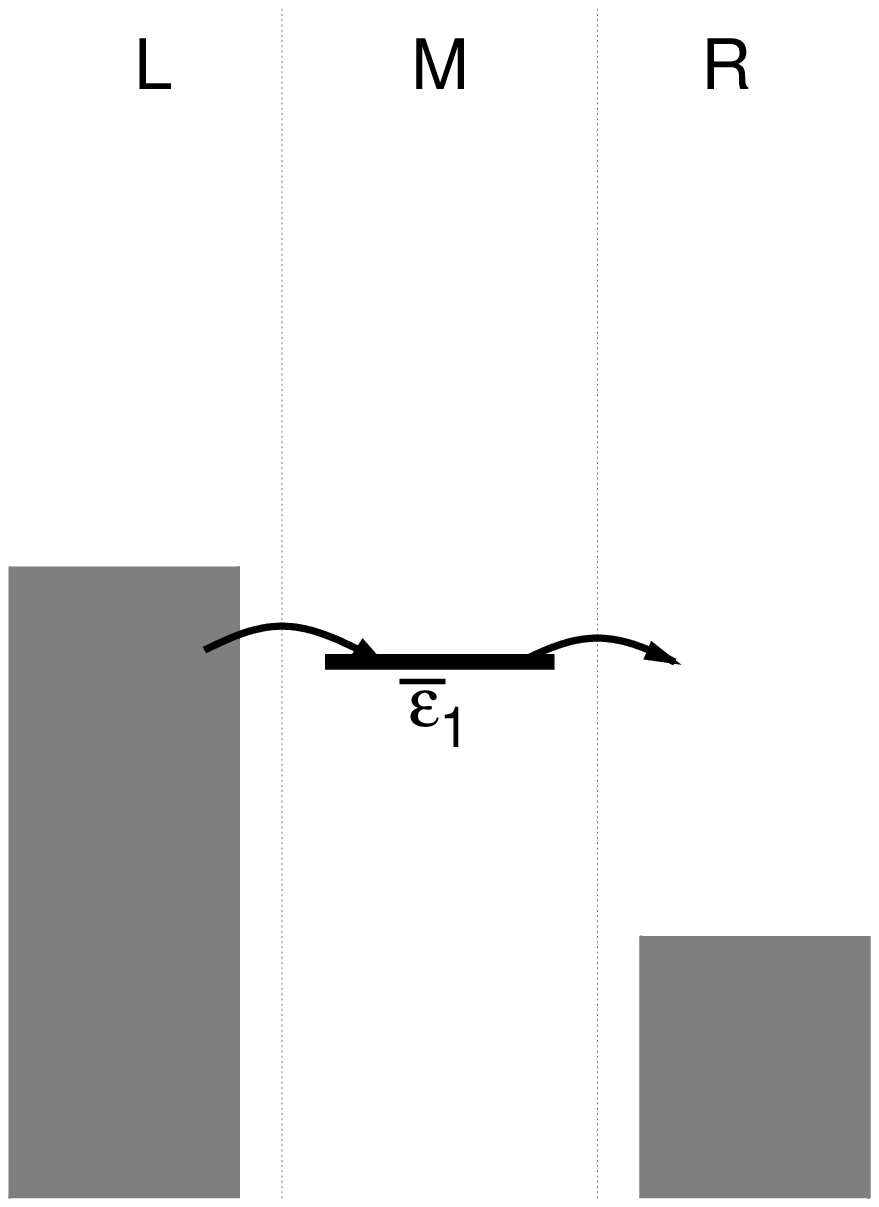}
}&
\resizebox{\newwidthprime}{\newheightprime}{
\includegraphics{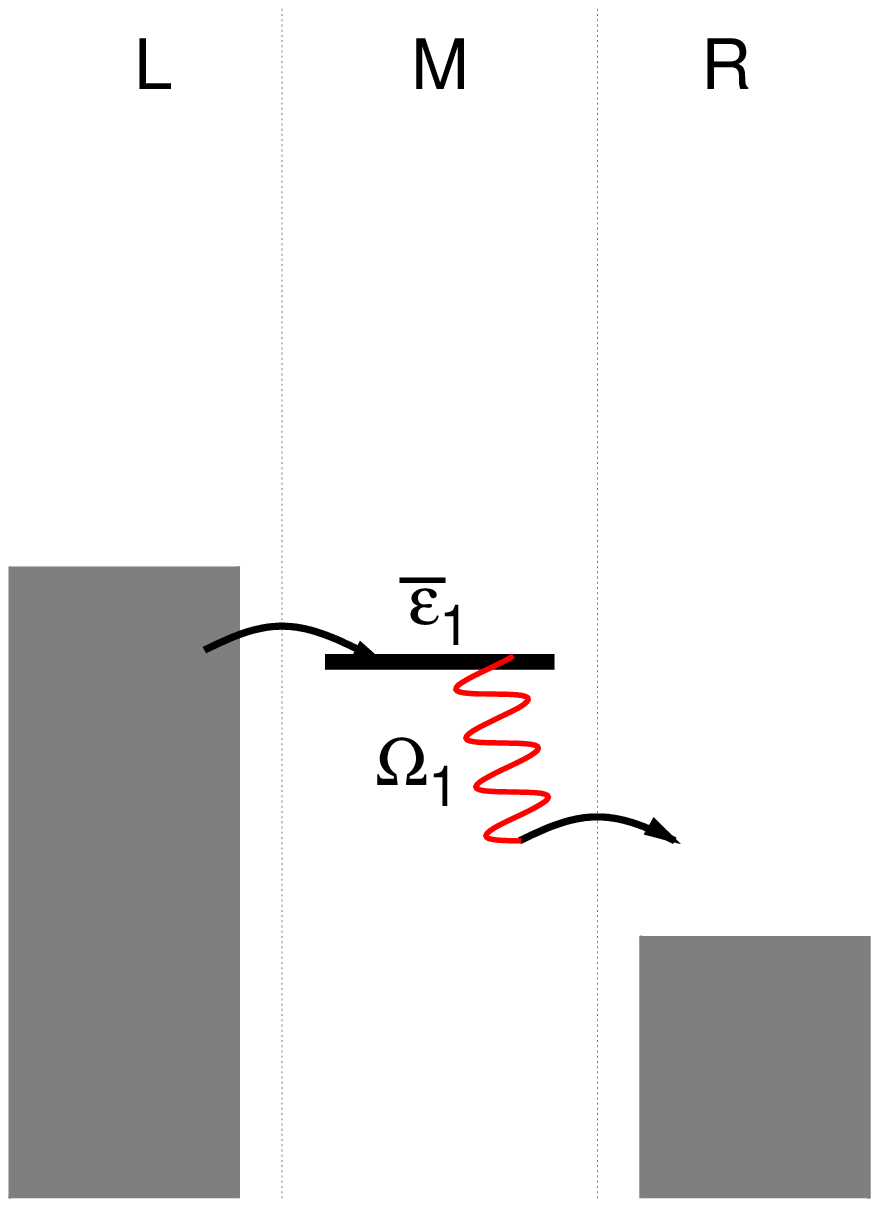}
}&
\resizebox{\newwidthprime}{\newheightprime}{
\includegraphics{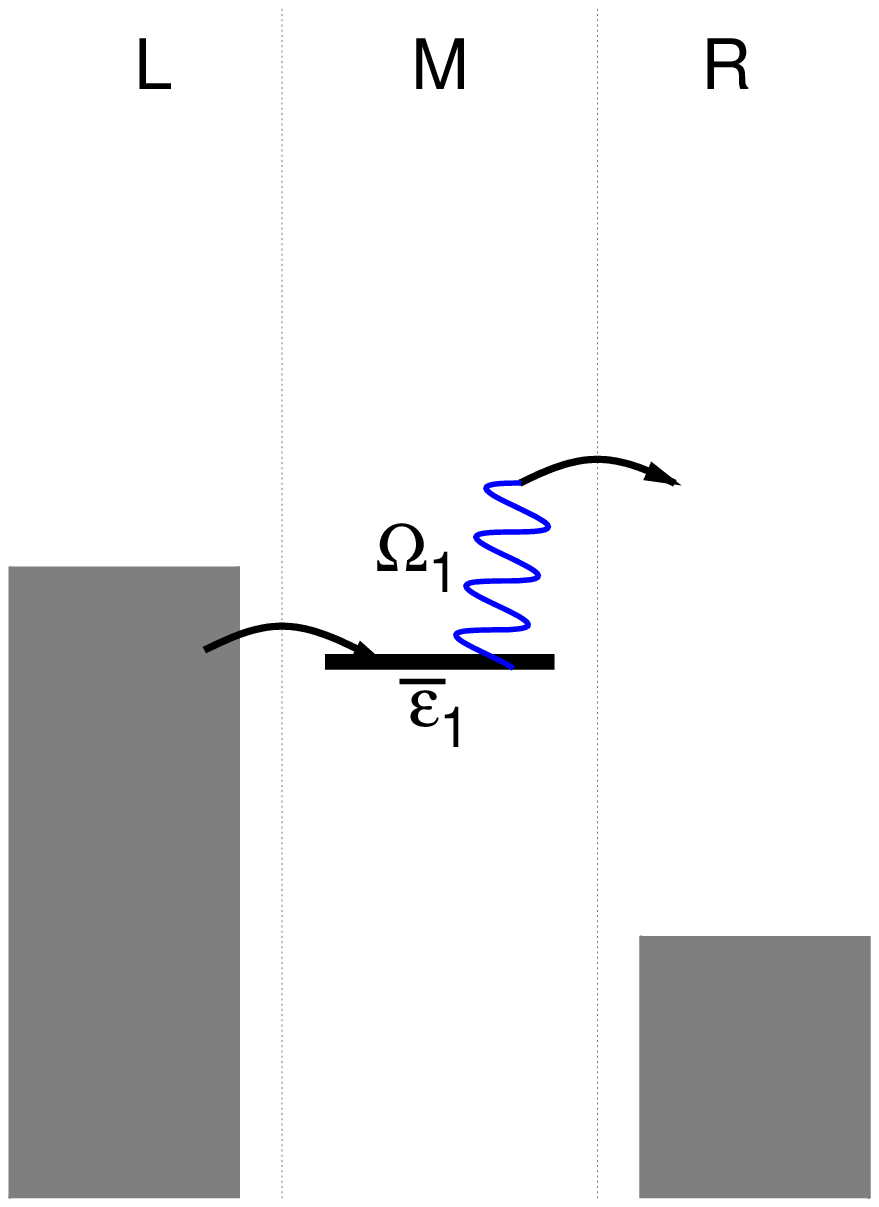}
}&
\resizebox{\newwidthprime}{\newheightprime}{
\includegraphics{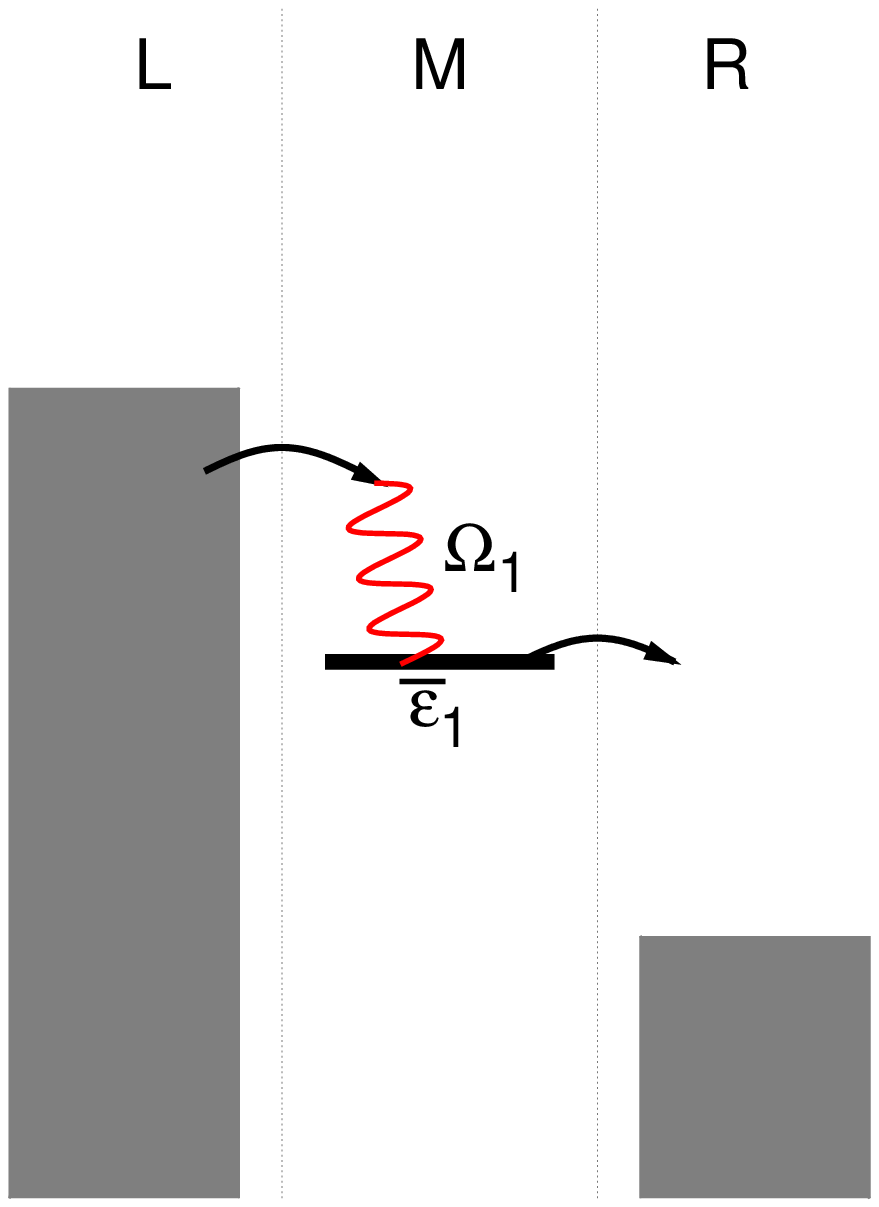}
}
\\ 
\end{tabular}
\end{center}
\caption{\label{basmech} Schematic representation of example processes for 
sequential tunneling in a molecular junction. 
Panel (a) shows sequential tunneling of an electron from the left lead to the right lead 
that involves two consecutive tunneling processes: 
from the left lead onto the molecular bridge and 
from the molecular bridge to the right lead. 
In panel (b)/(c) the latter of the two tunneling 
processes is accompanied by an excitation/deexcitation process, 
where due to electronic-vibrational coupling 
the vibrational mode is singly excited/deexcited. 
While the processes depicted by Panel (a) -- (c) become active 
at the same bias voltage, 
that is for $e\Phi\approx2\overline{\epsilon}_{1}$, 
resonant excitation processes like the one depicted by Panel (d) require 
higher bias voltages, 
$e\Phi\gtrsim2\overline{\epsilon}_{1}+\Omega_{1}$.}
\end{figure*}

Within the electronic scenario, the current-voltage characteristic displays a single step at $e\Phi=2\epsilon_{1}$, 
which is associated with the onset of resonant electronic transport processes where electrons are 
tunneling through the junction in two subsequent resonant tunneling events (see Fig.\ \ref{basmech}a). 
In contrast, the vibronic scenarios provide a number of additional resonant transport channels 
that become active at different bias voltages \cite{Hartle2010b}. Some of them are readily active at the onset of the resonant 
transport regime at $e\Phi=2\overline{\epsilon}_{1}$ and include both resonant excitation (Fig.\ \ref{basmech}b) 
and resonant deexcitation processes (Fig.\ \ref{basmech}c). 
Another set of resonant excitation processes (Fig.\ \ref{basmech}d) 
becomes active at higher bias voltages, $e\Phi=2(\overline{\epsilon}_{1}+n\Omega_{1})$ ($n\in\mathbb{N}$),   
giving rise to additional steps in the vibronic current-voltage characteristics.

The height of the steps in the current-voltage characteristics is different for all three transport scenarios, 
reflecting the fact that the corresponding transport processes strongly depend  
on the electronic-vibrational coupling strength $\lambda_{11}$ and the vibrational excitation level of the junction. 
The respective probabilities can be expressed in terms of the so-called Franck-Condon matrix elements $F_{mn}$, 
that is the transition matrix elements from the $m$th to the $n$th state of the vibrational mode upon change of the
charge state of the molecule. 
If, for example, electronic transport processes (Fig.\ \ref{basmech}a) occur with a relative probability of $1$ 
in the electronic scenario, they occur with a reduced probability of $F_{00}\approx0.7$ in the thermally equilibrated 
transport scenario. This value becomes even smaller in the vibronic transport scenario, $\sum_{n}p_{n}F_{nn}<F_{00}$, 
where $p_{n}$ represents the average population of the $n$th vibrational level. The first step 
in the current-voltage characteristics at $e\Phi=2\overline{\epsilon}_{1}$ is correlated with these 
probabilities, although a quantitative analysis 
is much more involved \cite{Hartle2010b}. The subsequent steps at $e\Phi=2(\overline{\epsilon}_{1}+n\Omega_{1})$ 
become successively smaller, which can be qualitatively 
understood by the reduced Franck-Condon matrix elements of the respective transport processes that involve 
transitions from the vibrational ground state to the $n$th 
excited state of the vibration, $F_{0n}=\text{e}^{-\lambda_{11}^{2}/\Omega_{1}^{2}} (\lambda_{11}^{2n}/\Omega_{1}^{2n})/n!$. 
However, as the comparison of the dashed black and the solid black line shows, 
vibrational nonequilibrium effects significantly modify the step structure and lead, in general, to a suppression 
of the electrical current that is flowing through junction E1V1 \cite{Hartle2010b}.

\begin{figure} 
\begin{center}
\resizebox{\newwidth}{\newheight}{
\includegraphics{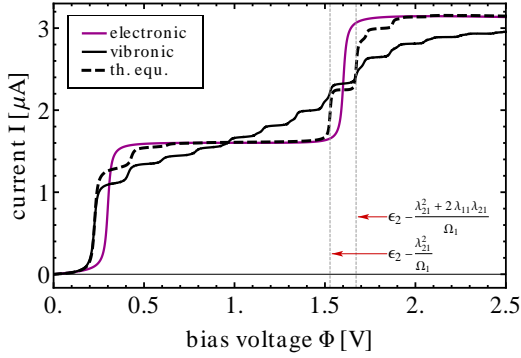}
}
\end{center}
  \caption{\label{ME44-thermal} 
Current-voltage characteristics 
of junction E2V1 (cf.\ Tab.\ \ref{tableII}). 
}
\end{figure}

So far, we have considered transport processes for a model with a single electronic state. 
In many cases, however, electron transport through single-molecule junctions is carried by 
a multitude of states. The resulting vibrational processes 
that may occur with respect to each of these states leads to 
a number of interesting (nonequilibrium) phenomena (see, e.g., Refs.\ \cite{Hartle09,Saito2009,Romano10,Hartle2010b,Hartle2012}). 
As an example, we consider 
a junction with two electronic states (model E2V1, see Tab.\ \ref{tableII} for a complete list of parameters). 
The corresponding current-voltage characteristics are shown in Fig.\ \ref{ME44-thermal}.

\begin{figure}
\begin{center}
\begin{tabular}{lll}
(a)&(b)&(c)
\\
\hspace{-0.25cm}\resizebox{\newwidthprime}{\newheightprime}{
\includegraphics{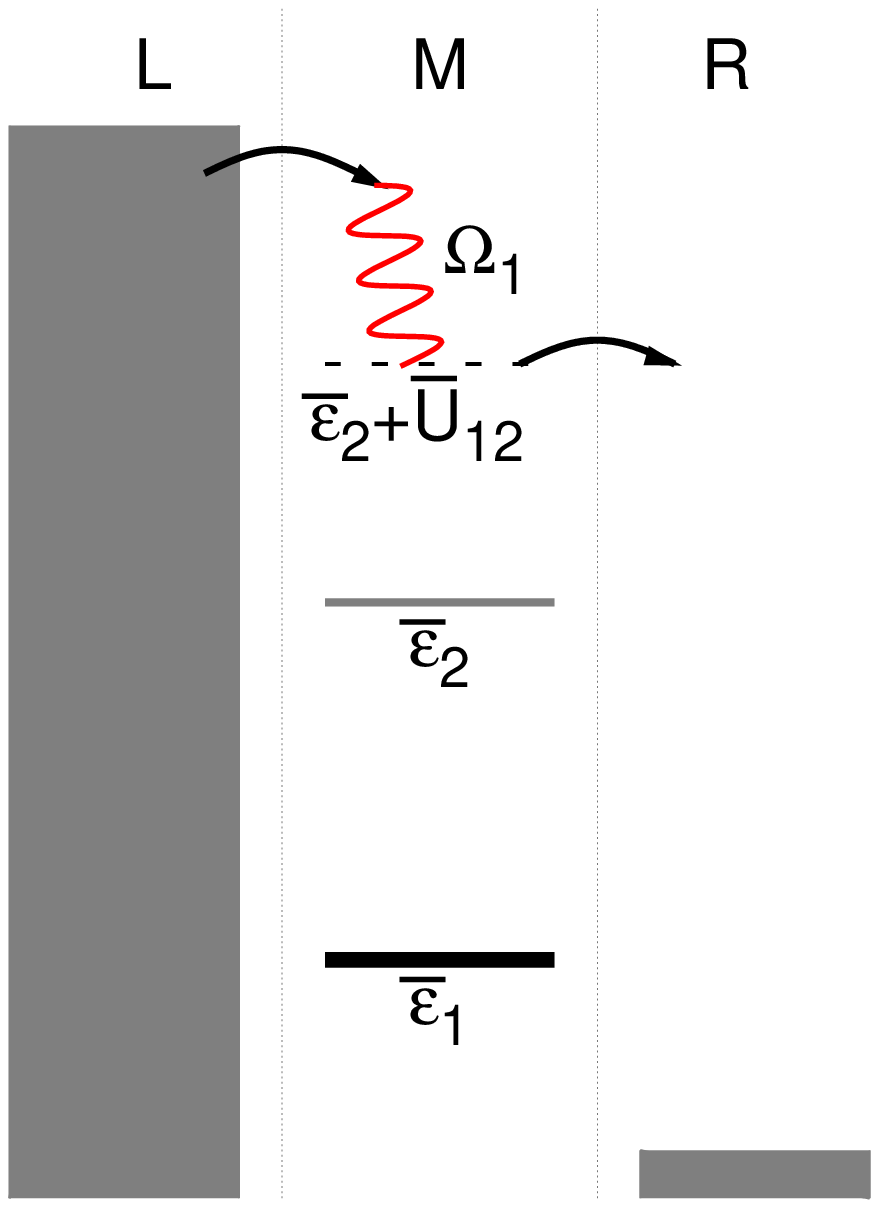}
}
&
\hspace{-0.1cm}\resizebox{\newwidthprime}{\newheightprime}{
\includegraphics{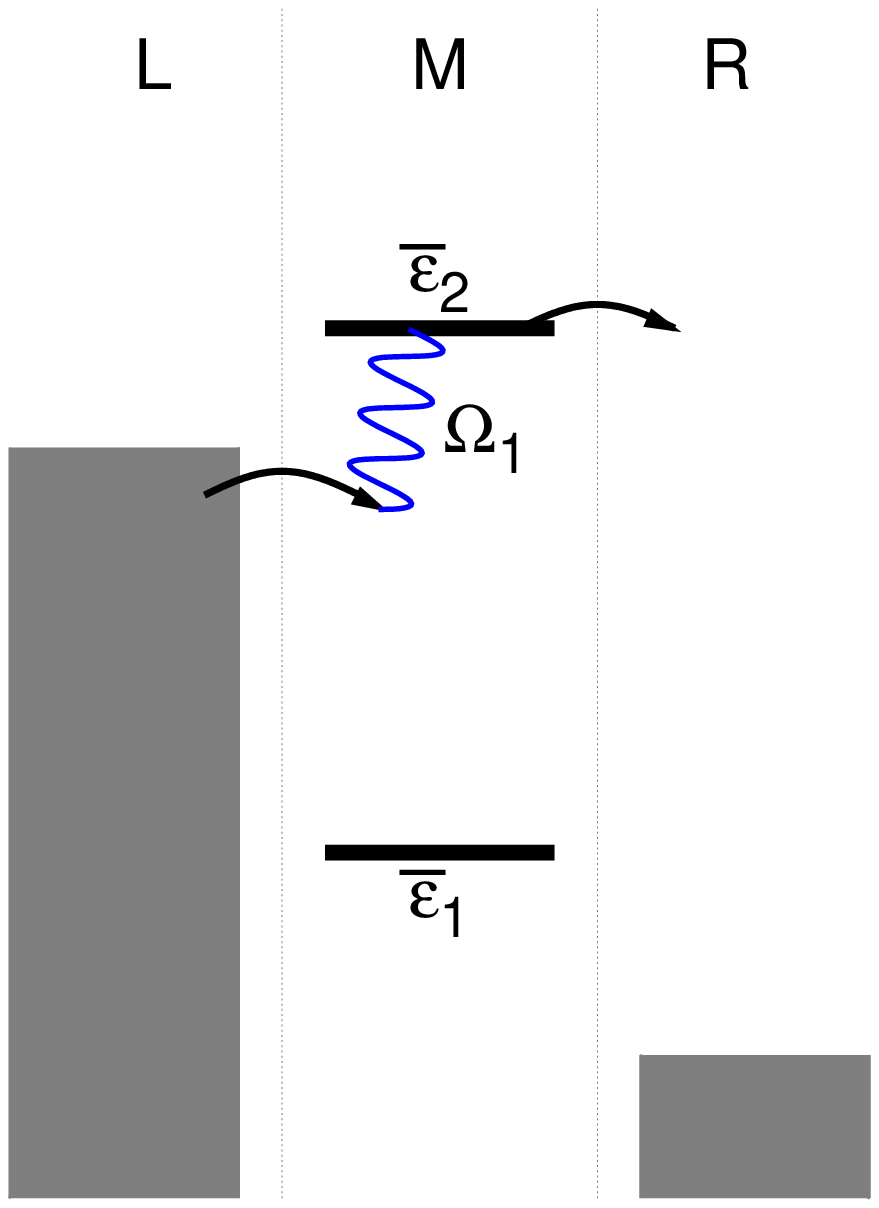}
}&
\hspace{-0.1cm}\resizebox{\newwidthprime}{\newheightprime}{
\includegraphics{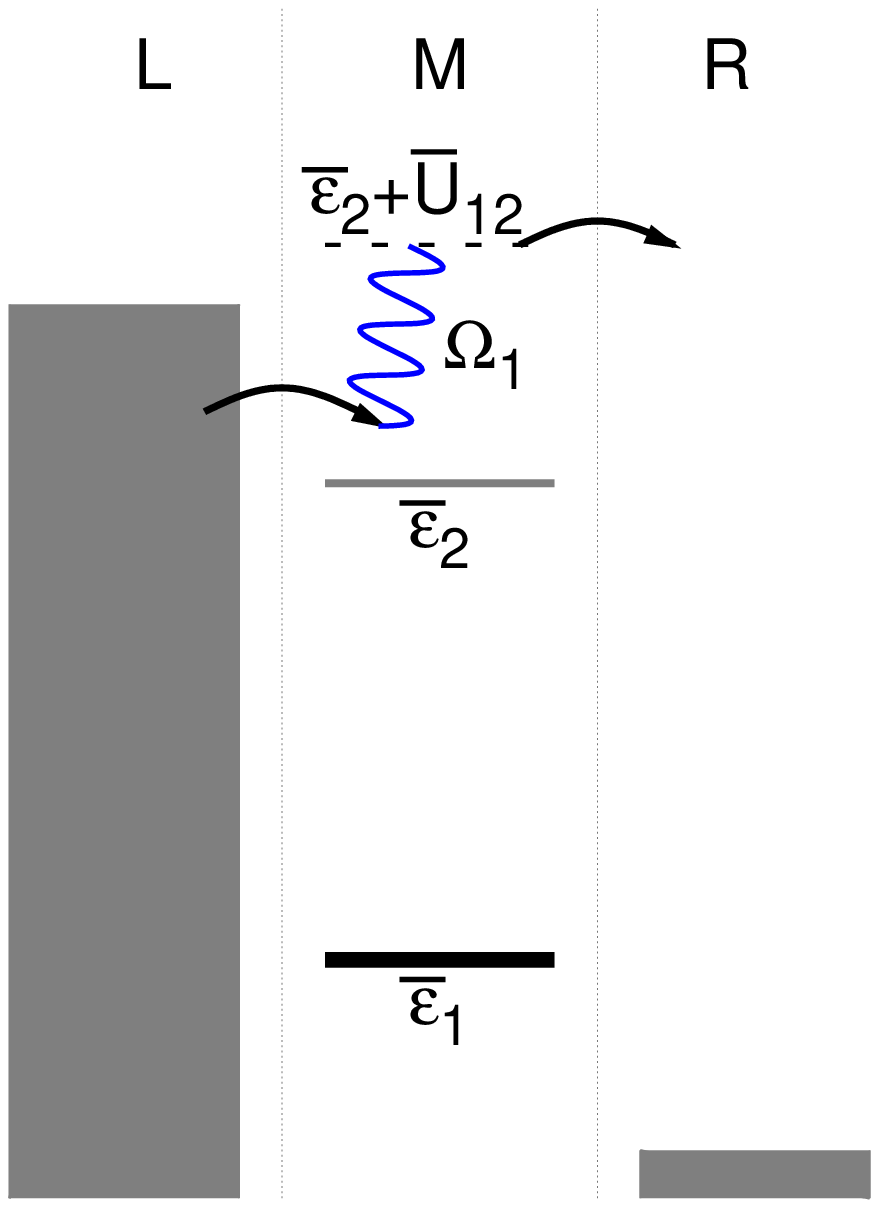}
}
\\ 
\end{tabular}
\end{center}
\caption{\label{twostateprocesses} Examples of sequential tunneling processes involving two electronic states. 
Panel (a) depicts a sequential tunneling process, 
which involves an excitation processes with respect to state 2. 
Thereby, 
it is assumed that state 1 is occupied, that is the tunneling electron requires an energy of 
$\epsilon_{2}+U_{12}+\Omega_{1}$, which is increased by the charging energy $U_{12}$   
due to interaction with the electron in state 1. 
Panels (c) and (d) show corresponding deexcitation processes, 
where the lower-lying electronic state is unoccupied and occupied, respectively. 
}
\end{figure}

The current-voltage characteristic associated with the electronic scenario shows two steps 
at $e\Phi=2\epsilon_{1}=0.3$\,eV and $e\Phi=2\epsilon_{2}=1.6$\,eV, 
indicating the onset of electronic transport processes through states $1$ and $2$, respectively. 
Similarly, the thermally equilibrated transport shows three pronounced steps 
at $e\Phi=2\overline{\epsilon}_{1}=0.228$\,eV, $e\Phi=2\overline{\epsilon}_{2}=1.528$\,eV and 
$e\Phi=2(\overline{\epsilon}_{2}+\overline{U}_{12})=1.672$\,eV, where the first of these steps 
is, just as in the electronic scenario, associated with the onset of resonant electronic transport 
processes through the first electronic level. 
The other two steps originate from the onset of resonant electronic transport processes through the second electronic state. 
However, due to electron-electron interactions, $\overline{U}_{12}=72$\,meV, 
transport through this state depends on wether the 
first level is occupied or unoccupied. This leads to a splitting of the corresponding step into two steps that are separated 
by $e\Delta\Phi=2\overline{U}_{12}$. 
Apart from the splitting of steps due to electron-electron interactions, 
the current-voltage characteristics of the thermally equilibrated scenario can be understood 
on the same grounds as for a junction with a single electronic level. 
Vibrational side-steps are observed with respect to each of the three electronic resonances, 
that is at $e\Phi=2(\overline{\epsilon}_{1}+n\Omega_{1})$, 
$e\Phi=2(\overline{\epsilon}_{2}+n\Omega_{1})$ and $e\Phi=2(\overline{\epsilon}_{2}+\overline{U}_{12}+n\Omega_{1})$ ($n\in\mathbb{N}$). 
They indicate, as before, the onset of resonant excitation processes (see Figs.\ 
\ref{basmech}d and \ref{twostateprocesses}a). 
This shows that the onset of vibrational processes does not only depend on the position of the electronic levels, 
given by $\overline{\epsilon}_{m}$, but also on the electron-electron interaction strengths $\overline{U}_{mn}$.

In contrast, the vibronic scenario shows a more complex behavior. 
For example, it exhibits only a single pronounced step 
at $e\Phi=2\overline{\epsilon}_{1}$. The two steps associated with the onset of resonant transport 
through the second electronic state are barely visible. 
This is due to the current-induced level of vibrational excitation, which is generated by 
inelastic processes with respect to the first state (Fig.\ \ref{basmech}d). This 
leads, on one hand, to a reorganization of the steps heights and an overall reduction of 
the current level through state 1 (cf.\ the discussion of Fig.\ \ref{K21-current} 
or see Ref.\ \cite{Hartle2010b} for a more detailed analysis of these phenomenona). 
On the other hand, it triggers resonant deexcitation 
processes with respect to the second electronic state (Figs.\ \ref{twostateprocesses}b and \ref{twostateprocesses}c)  
even before it has entered the bias window \cite{Hartle09,Hartle2010b}. 
As a result, the steps corresponding to transport through state 2 are effectively broadened such that, 
instead of two pronounced steps, the onset of resonant transport through the second electronic state is 
associated with a number of small steps at $e\Phi=2(\overline{\epsilon}_{2}-n\Omega_{1})$ 
and $e\Phi=2(\overline{\epsilon}_{2}+\overline{U}_{12}-n\Omega_{1})$ \cite{Hartle09}. 
This shows that vibrational nonequilibrium effects have a pronounced influence on the electrical transport properties of 
a molecular junction and that a more detailed and quantitative understanding of these effects 
requires an analysis of the corresponding vibrational excitation characteristic 
(which will be given in Sec.\ \ref{SecElHole}).

\subsection{Electron-hole pair creation processes} 
\label{SecElHole}

\begin{figure}
\begin{center}
\begin{tabular}{ll}
(a)&(b) 
\\
\resizebox{\newwidthprime}{\newheightprime}{
\includegraphics{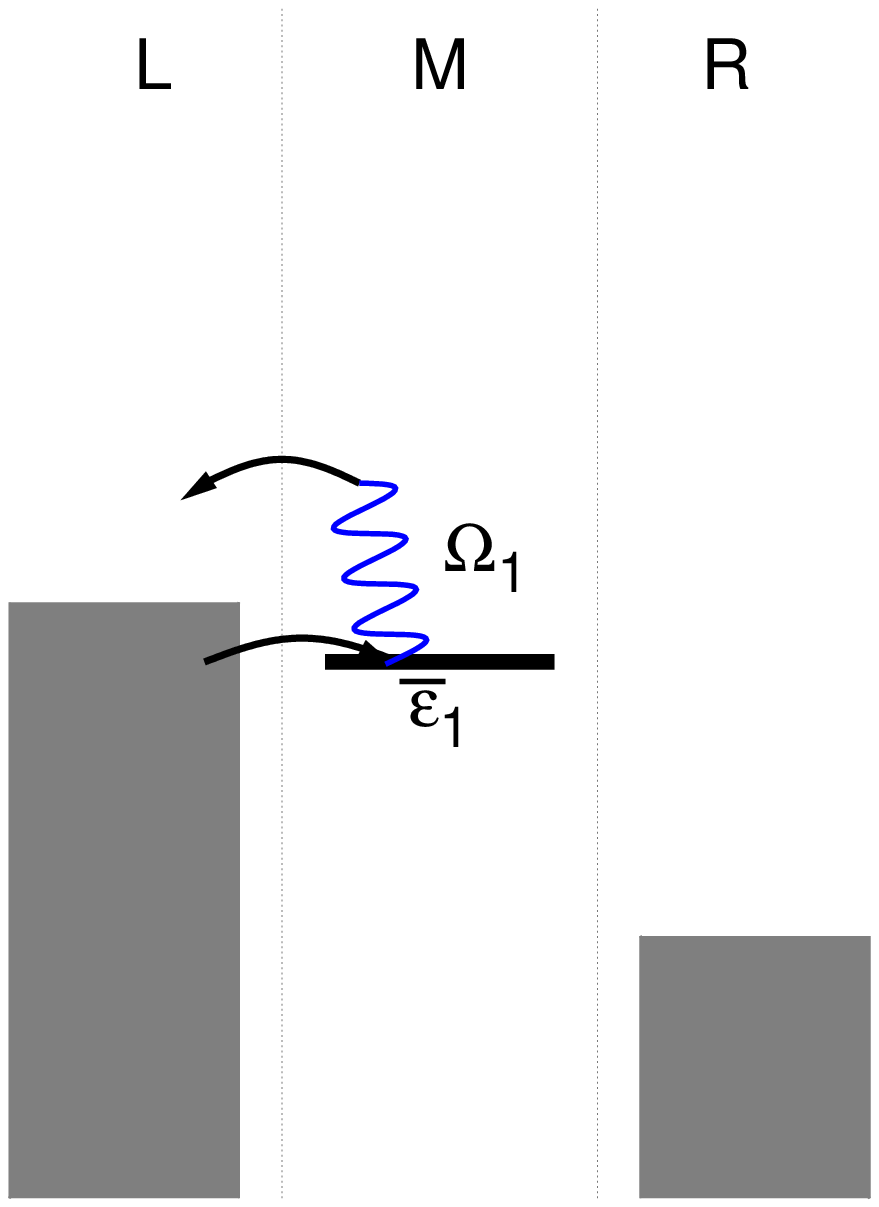}
}
&
\resizebox{\newwidthprime}{\newheightprime}{
\includegraphics{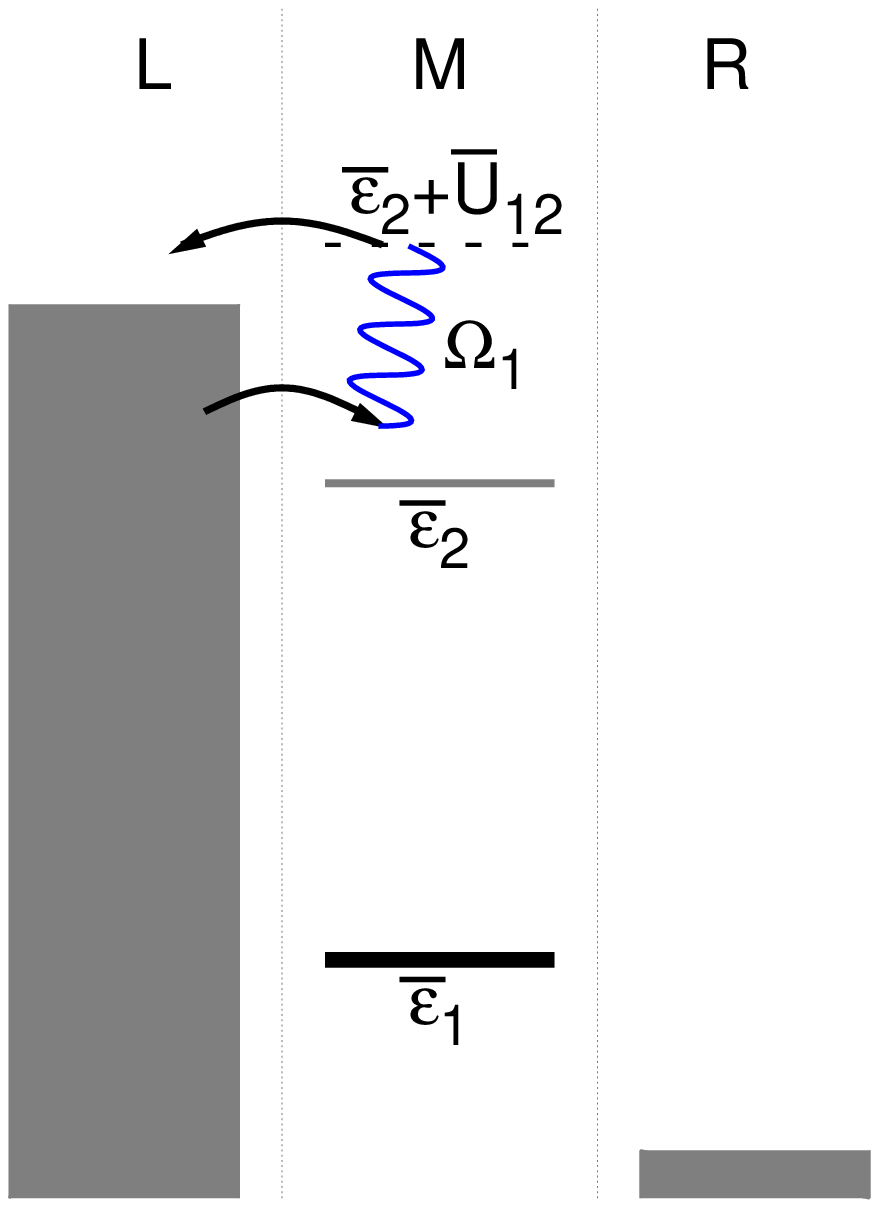}
}
\\ 
\end{tabular}
\end{center}
\caption{\label{el-h-pair-creation} Examples of 
electron-hole pair creation processes. 
Panel (a) shows a resonant electron-hole pair creation process, 
where in two sequential tunneling events an electron 
tunnels from the left lead onto 
the molecular bridge and back again to the left lead. 
Thereby, the electron takes up a quantum of vibrational energy (blue wiggly line). 
A pair creation process with respect to a higher-lying electronic state is 
shown in Panel (b). 
}
\end{figure}

In this section, we analyze the vibrational excitation characteristics of junctions E1V1 and E2V2. 
This requires to account for all processes, where the molecule exchanges energy with the electrodes, 
including inelastic transport processes (see Figs.\ \ref{basmech} and \ref{twostateprocesses}) as well as 
electron-hole pair creation processes (examples of which are depicted in Fig.\ \ref{el-h-pair-creation}). 
Pair creation and transport processes involve the same tunneling processes and, therefore, occur principally 
with the same probability. They are distinguished only by the fact that the tunneling electrons 
return to the same lead in the course of a pair creation event, 
while they transfer from one lead to the other in the course of a transport process.

\begin{figure} 
\begin{center}
\resizebox{\newwidth}{\newheight}{
\includegraphics{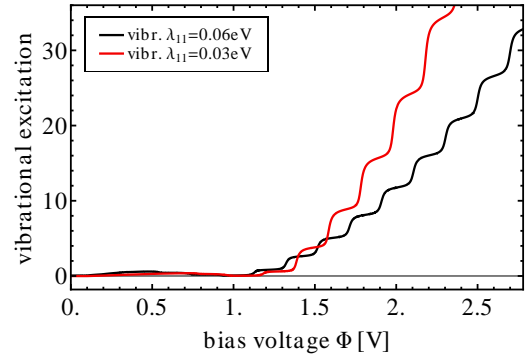}
}
\end{center}
  \caption{\label{K21-nBocc} Vibrational excitation characteristic of junction 
junction E1V1 (cf.\ Tab.\ \ref{tableI}). }
\end{figure}

The excitation characteristics of junction E1V1 is depicted by the solid black line in Fig.\ \ref{K21-nBocc}. 
It shows 
a qualitatively different behavior with respect to the applied bias voltage $\Phi$ than the corresponding 
current-voltage characteristics (cf.\ Sec.\ \ref{SecBas}). 
Although it exhibits steps at the same bias voltages, 
the height of these steps follows clearly another phenomenology. 
While the steps in the current-voltage characteristic become successively smaller with an increasing bias voltage $\Phi$, 
they become successively larger in the vibrational excitation characteristic. 
Thus, in contrast to the current-voltage characteristics, the level of vibrational 
excitation does not saturate for high bias voltages \cite{Hartle2011}. 
Moreover, the step heights increase  
if the electronic-vibrational coupling strength is reduced. This can be inferred from the comparison 
of the vibrational excitation characteristic of junction E1V1 with the excitation characteristic of a very similar 
junction that differs from junction E1V1 only by a reduced electronic-vibrational coupling strength $\lambda_{11}=0.03$\,eV. 
It is depicted by the solid red line in Fig.\ \ref{K21-nBocc}.  
This behavior 
is rather counter-intuitive, because the transition matrix elements for the 
corresponding excitation processes (Fig.\ \ref{basmech}d) become smaller 
for weaker electronic-vibrational coupling strengths $\lambda_{11}$ \cite{Mitra04,Semmelhack,Avriller2010,Hartle2010b,Hartle2011}.

To understand the phenomenology of the vibrational excitation characteristics  
it is expedient to consider the nature of this observable. 
In contrast to the electrical current, which is determined by the number of transport processes in a given time interval, 
the level of vibrational excitation (in steady state) is determined by the ratio of the probabilities for excitation and deexcitation 
processes. 
At low bias voltages, $e\Phi\lesssim2\overline{\epsilon}_{1}$, 
vibrational excitation can only be generated in tunneling processes with respect to the right lead (cf.\ Fig.\ \ref{basmech}b). 
Resonant deexcitation, however, is less restricted and can occur with respect to both leads (\emph{e.g.}\ 
by the processes that are depicted in Figs.\ \ref{basmech}c and \ref{el-h-pair-creation}a). This leads, roughly speaking, 
to a ratio between the probabilities for excitation and deexcitation of the vibrational mode 
of at least one to two. 
The corresponding level of vibrational excitation is, therefore, rather low 
and originates, to a large extent, from 
polaron formation (see Eqs.\ (\ref{formulavibex}) and (\ref{formulavibex2})). 
At higher bias voltages $e\Phi > 2(\overline{\epsilon}_{1}+\Omega_{1})$, however, 
the imbalance between excitation and deexcitation processes becomes gradually smaller. 
This is, on one hand, due to an increasing number of excitation processes that become active 
but, on the other hand, also due to a decreasing number of deexcitation processes 
that were active at smaller bias voltages. 
For example, the electron-hole pair creation process 
depicted in Fig.\ \ref{el-h-pair-creation}a is active at the onset of the resonant transport regime but is blocked 
for bias voltages $e\Phi>2(\overline{\epsilon}_{1}+\Omega_{1})$. 
Thus, the probabilities for excitation and deexcitation of the vibrational mode 
become successively the same, resulting in an indefinite increase of 
vibrational excitation at higher bias voltages, where only transport processes occur \cite{Hartle2011}. 
Moreover, as the pair-creation processes that are blocked at $e\Phi>2(\overline{\epsilon}_{1}+n\Omega_{1})$ 
are more important for weaker electronic-vibrational coupling, 
the corresponding increase/step in the vibrational excitation characteristic becomes larger for 
smaller values of $\lambda_{11}$ \cite{Hartle2010b,Hartle2011}.

\begin{figure}
\begin{center}
\resizebox{\newwidth}{\newheight}{
\includegraphics{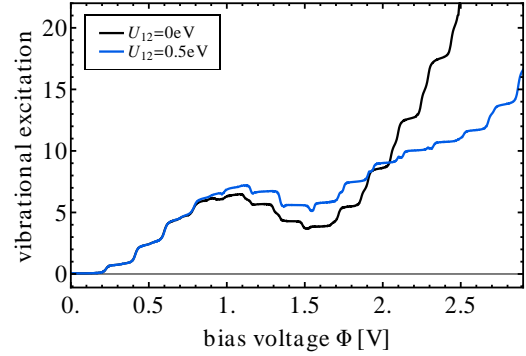}
}
\end{center}
  \caption{\label{FigCoulCool} Vibrational excitation characteristics for junction E2V1 
  (solid black line, cf.\ Tab.\ \ref{tableII}). In addition, 
  the vibrational excitation characteristics of a very similar system is shown, which, in contrast to junction E2V1, 
  includes additional repulsive electron-electron interactions, $U_{12}=0.5$\,eV (solid blue line). 
}
\end{figure}

Similar as in Sec.\ \ref{SecBas}, we extend our considerations at this point 
to a junction with two electronic states, model E2V1. 
The vibrational excitation characteristic of this junction is shown in Fig.\ \ref{FigCoulCool}.  
As long as the second electronic state is located far outside the bias window (for $\Phi\lesssim1$\,V), 
the level of vibrational excitation increases monotoneously due to resonant 
excitation processes with respect to the first electronic state (Fig.\ \ref{basmech}d). 
At higher bias voltages, however, the vibrational energy thus generated 
facilitates resonant deexcitation processes with respect to the second electronic state, 
including transport (Figs.\ \ref{twostateprocesses}b and \ref{twostateprocesses}c) 
as well as electron-hole pair creation processes (Figs.\ \ref{el-h-pair-creation}b). 
These processes significantly reduce the level of vibrational excitation before the second electronic state enters the bias window 
(\emph{i.e.} $1$\,eV$<e\Phi<2\overline{\epsilon}_{2}$). The net effect is that the presence of the second electronic state 
results in a cooling of the junction at these bias voltages \cite{Hartle09,Hartle2010b}.

This cooling extends to even higher bias voltages, if electron-electron interactions are present. 
This can be seen by comparison of the solid black line in Fig.\ \ref{FigCoulCool} with 
the solid blue line, which shows the vibrational excitation characteristic 
of a junction that differs from junction E2V1 only by additional 
repulsive electron-electron interactions, $U_{12}=0.5$\,eV. 
Due to the more pronounced splitting of the resonances that are associated with the second electronic state in this system, 
deexcitation processes due to both transport (Fig.\ \ref{twostateprocesses}c) 
and electron-hole pair creation processes (Fig.\ \ref{el-h-pair-creation}b) are shifted to higher energies and, consequently,  
become active at higher bias voltages. Thus, the additional electron-electron interactions lead to a slight increase of 
the level of vibrational excitation at low bias voltages (\emph{i.e.}\  for $\Phi\lesssim2$\,V). 
At higher bias voltages, however, they lead to a significant decrease of the vibrational excitation level 
(Coulomb Cooling) \cite{Hartle2010b}.

\subsection{Transport phenomena due to electron-hole pair creation processes}
\label{transphen}

The results discussed in the preceding section, Sec.\ \ref{SecElHole}, 
show that a detailed understanding of the vibrational excitation characteristic of a single-molecule contact 
can only be obtained if electron-hole pair creation processes are taken into account. 
Thus, pair creation processes also influence 
the corresponding electrical transport properties, as the efficiency of transport processes is strongly 
interrelated with the vibrational excitation levels (cf., \emph{e.g.}, 
the reduced current level in the resonant transport regime of junctions E1V1 and E1V2  
discussed in Sec.\ \ref{SecBas}). 
In this section, we show that this influence may be substantial and that it can lead to a number of interesting 
transport phenomena such as, for example, negative differential resistance (Sec.\ \ref{ndrsec}), rectification (Sec.\ \ref{vibrec}), 
mode-selective vibrational excitation (Sec.\ \ref{msve})
or a pronounced temperature dependence of the current 
in the presence of destructive quantum interference effects (Sec.\ \ref{tempdep}).

\subsubsection{Negative differential resistance}
\label{ndrsec}

\begin{figure} 
\begin{center}
\resizebox{\newwidth}{\newheight}{
\includegraphics{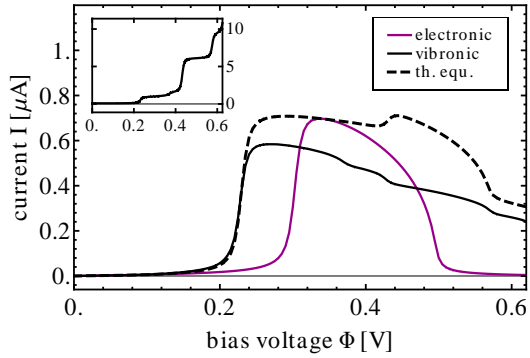}
}
\end{center}
\caption{\label{NDR-CP5} Current-voltage characteristics of a molecular junction  
that exhibits negative differential resistance (model BAND, cf.\ Tab.\ \ref{tableI}). 
The inset shows the corresponding vibrational excitation characteristic. 
} 
\end{figure}

In general, an electronic device exhibits negative differential resistance (NDR) when  
its differential conductance $\text{d}I/\text{d}\Phi$ becomes negative. 
This phenomenon is revelant for a number 
of applications in electronics, for example, in analog-digital converters \cite{Broekart1998}  
or logic circuits \cite{Mazumder1998,Mathews1999}. 
A prime example for a nanoelectronic device that exhibits NDR is the resonant 
tunneling diode, where the overlap of the (narrow) conduction bands in the leads 
decreases with increasing bias voltage $\Phi$ \cite{Davies93,Hyldgaard94}. 
The current-voltage characteristic of a corresponding model for a molecular junction (model BAND, cf.\ Tab.\ \ref{tableI}) 
is depicted by the solid purple line in Fig.\ \ref{NDR-CP5}. 
At low bias voltages, where the conduction bands overlap with each other, 
the current level of the junction increases, in particular when resonant transport processes through 
the electronic state become active at $e\Phi\approx2\epsilon_{1}$. At higher bias voltage, however, 
the overlap of the conduction bands continuously decreases such that electronic transport processes 
(Fig.\ \ref{basmech}a) can no longer connect an occupied state in one of the leads with an 
unoccupied state in the other lead. Thus, the current level of the junction decreases.

In the presence of electronic-vibrational coupling, the effect of the narrow conduction bands is significantly reduced, 
as can be seen by the solid black line in Fig.\ \ref{NDR-CP5}. This is due to 
inelastic transport processes, which allow electrons to span a much wider range of energies in the course of a transport process, 
increasing effectively the overlap of the conduction bands. 
In addition, however, another much more narrow NDR feature occurs at $e\Phi=2(\overline{\epsilon}_{1}+\Omega_{1})$. 
It is in line with an increase of the vibrational excitation level of the junction 
(depicted in the inset of Fig.\ \ref{NDR-CP5}). The thermally equilibrated 
scenario (dashed black line), however, does not exhibit NDR at this bias voltage. On the contrary, it displays 
an increase of the current level. Thus, we trace this NDR effect back to the suppression of 
electron-hole pair creation processes (Fig.\ \ref{el-h-pair-creation}a) 
and the associated increase in vibrational excitation 
rather than the opening of another inelastic channel (Fig.\ \ref{basmech}d), although both 
occur at the same bias voltage. Note that this NDR effect 
is not related to the narrow conduction bands of this model system 
and that it may be more pronounced if the molecule is asymmetrically coupled to the leads 
(see, for example, Refs.\ \cite{Schoeller01,Zazunov06,Hartle2010b}).

\subsubsection{Vibrational rectification and spectroscopy in molecular junctions 
}
\label{vibrec}

\begin{figure} 
\begin{center}
\resizebox{\newwidth}{\newheight}{
\includegraphics{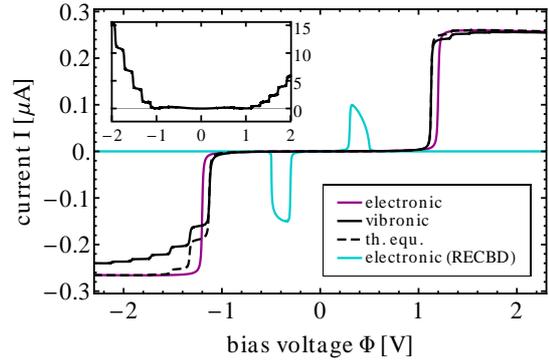}
}
\end{center}
  \caption{\label{ElRec} Current-voltage characteristics of the asymmetric 
junction REC (cf.\ Tab.\ \ref{tableI}). 
The inset shows the corresponding vibrational excitation characteristic. 
The solid turquoise line refers to model RECBD (cf.\ Tab.\ \ref{tableI}), 
which is distinguished from junction REC in particular due to a much narrower band width $\gamma$. 
}
\end{figure}

Diode-like behavior 
represents another important device characteristic \cite{Elbing05}. 
Such behavior is often associated with an asymmetric coupling of the molecule to the leads. 
But an asymmetry in the molecule-lead coupling is not sufficient \cite{Mujica2002}, 
as is demonstrated by the solid purple line in Fig.\ \ref{ElRec}. 
It shows the electronic current-voltage characteristics of 
junction E1V1 with a reduced coupling stength to the right lead (model REC, cf.\ Tab.\ \ref{tableI}). 
It is almost perfectly antisymmetric with respect to the polarity of the bias voltage, 
\emph{i.e.}\ $I(\Phi)=-I(-\Phi)$. 
The origin of this behavior is twofold. First, 
transport processes involve the coupling to both leads and, therefore, cannot transfer 
an asymmetry in the coupling to the leads to the transport characteristics of a molecular contact. 
Second, the density of states for these processes does also not display a pronounced asymmetry with respect to 
the bias voltage $\Phi$. In particular in the limit $\nu_{\text{R},1}/\nu_{\text{L},1}\rightarrow0$, 
where tunneling processes with respect to the right lead represent the bottleneck for transport, 
it is given by $\Gamma_{\text{R},11}(\epsilon_{1})/\nu_{\text{R},1}^{2}$. 
If, however, in addition to an asymmetric molecule-lead coupling, 
the band width in the leads is reduced (from $\gamma=3$\,eV to $\gamma=0.2$\,eV), 
the density of states $\Gamma_{\text{R},11}(\epsilon_{1})/\nu_{\text{R},1}^{2}$ 
is much smaller in the resonant transport regime for negative bias voltages ($e\Phi\lesssim2\overline{\epsilon}_{1}$) 
than it is for positive bias voltages ($e\Phi\gtrsim2\overline{\epsilon}_{1}$). 
In that case, the corresponding current-voltage characteristic exhibits an asymmetry with respect to the polarity of 
the applied bias voltage, which can be seen by the solid turquoise line in Fig.\ \ref{ElRec}, which shows the current-voltage 
characteristic of such a junction (model RECBD, cf.\ Tab.\ \ref{tableI}) 
\footnote{Note that for similar systems with $\epsilon_{1}\approx0$, the current response would be perfectly antisymmetric irrespective 
of an asymmetric molecule-lead coupling and the value of the band width $\gamma$.}.

Another mechanism that leads to a different current response at different polarities of the applied bias voltage 
is electronic-vibrational coupling \cite{Hartle2010b,Volkovich2011b}. This can be seen by the solid black 
and dashed black lines in Fig.\ \ref{ElRec}, 
which show the current-voltage characteristics of junction REC considering the vibronic and the thermally equilibrated 
transport scenario, respectively. While the black line exhibits a pronounced asymmetry of the current level, 
the dashed black line shows an asymmetry in the current level only in the vicinity $e\Phi\sim\pm2\overline{\epsilon}_{1}$. 
For the thermally equilibrated scenario, this behavior can be deduced again from the density of states for 
resonant transport processes in the right lead. For positive bias voltages, $e\Phi\gtrsim2\overline{\epsilon}_{1}$, 
it is given by $\sum_{n} F_{0n} \Gamma_{\text{R},11}(\epsilon_{1})/\nu_{\text{R},1}^{2}\approx \Gamma_{\text{R},11}(\epsilon_{1})/\nu_{\text{R},1}^{2}$, 
where the sum over the Franck-Condon matrix elements $F_{0n}$ indicates that electronic (Fig.\ \ref{basmech}a) 
as well as resonant excitation processes (Fig.\ \ref{basmech}b) become simultaneously active. For negative bias voltages, 
however, only electronic processes are active at $e\Phi=-2\overline{\epsilon}_{1}$, while excitation processes 
with respect to tunneling from the right lead become successively active at $e\Phi=-2(\overline{\epsilon}_{1}+n\Omega_{1})$. 
This leads to relative step heights in the current-voltage characteristics that are given by 
$F_{00}$, $F_{00}+F_{01}$, etc..

Although the same effects do also contribute to the asymmetric current response of the 
vibronic transport scenario, its behavior is strongly linked to the corresponding level of vibrational 
excitation (shown in the inset of Fig.\ \ref{ElRec}). As the excitation levels of the junction are much 
higher for negative bias voltages, the current is more strongly suppressed than for positive bias voltages. 
This relation between high levels of vibrational excitation 
and a reduction of the current level has already been outlined 
in Sec.\ \ref{SecBas} (and in Refs.\ \cite{Hartle2010b,Volkovich2011b}). 
As a consequence, the current suppression or the rectification effect also extends in the vibronic scenario 
over a wider range of bias voltages than for the thermally equilibrated scenario. 
Thereby, the asymmetry in vibrational excitation is due to electron-hole pair creation processes. 
In contrast to transport processes, they involve 
only one of the leads. Thus, in the limit $\nu_{\text{R},1}/\nu_{\text{L},1}\rightarrow0$, 
electron-hole pair creation processes with respect to the left lead (Fig.\ \ref{el-h-pair-creation}a) 
are much more important than pair creation processes with respect to the right lead as well as 
excitation and deexcitation processes due to transport processes. Therefore, the vibrational mode is more efficiently cooled 
for positive than for negative bias voltages, since pair creation processes with respect to the left lead 
involve less vibrational quanta and, therefore, are more effective for positive 
than for negative bias voltages.

\begin{figure} 
\begin{center}
\resizebox{\newwidth}{\newheight}{
\includegraphics{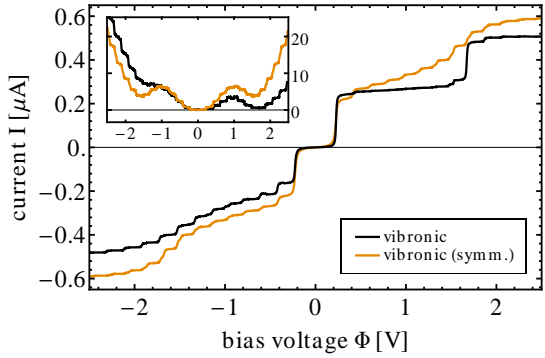}
}
\end{center}
  \caption{\label{SpecMolOrb} Current-voltage characteristics of the asymmetric 
junction SPEC (cf.\ Tab.\ \ref{tableII}). 
The inset shows the corresponding vibrational excitation characteristic. 
For comparison, the solid orange lines show results of the 
corresponding symmetric molecular junction E2V1 rescaled by a factor of $1/5$.  
}
\end{figure}

Considering spectroscopic applications of single-molecule junctions, 
the above described vibrational rectification effect leads to important implications. 
For example, one may observe a different number of side-steps (or different relative step heights) in 
the current-voltage characteristic of an asymmetrically coupled molecular junction 
by changing the polarity of the applied bias voltage 
(cf.\ Fig.\ \ref{ElRec}) \cite{Secker2010}. Furthermore, an asymmetry in the molecule-lead coupling may be useful 
to observe signals of states that are located 
further away from the Fermi level. According to our discussion of Fig.\ \ref{ME44-thermal} (Sec.\ \ref{SecBas}), 
these signals may be blurred by the effect of resonant deexcitation processes \cite{Hartle09}. 
In asymmetric junctions, however, where the effect of local cooling by electron-hole pair creation processes 
is different for different bias polarities, these processes can be less pronounced, at least  
for one polarity of the applied bias voltage \cite{HartlePhD}. This is shown in Fig.\ \ref{SpecMolOrb}, where the current-voltage 
characteristic of junction E2V1 is compared to a very similar junction that differs from junction E2V1 only by 
reduced coupling strengths to the right lead (model SPEC, cf.\ Tab.\ \ref{tableI}). In contrast to the symmetric junction E2V1, 
junction SPEC exhibits, due to the effect of a more pronounced cooling by electron-hole pair creation processes, 
two pronounced steps at $e\Phi=2\overline{\epsilon}_{1}$ and 
$e\Phi=2(\overline{\epsilon}_{2}+U_{12})$ that correspond to the onset of resonant transport processes 
through the first and the second electronic state.

\subsubsection{Mode-selective vibrational excitation}
\label{msve}

As our discussion of vibrational rectification effects in Sec.\ \ref{vibrec} showed, 
the effect of cooling by electron-hole pair creation processes can be controlled by the external bias voltage $\Phi$ 
if the molecule is asymmetrically coupled to the leads. Thereby, we focused on the electrical transport properties. 
In this section, we investigate junctions with multiple vibrational degrees of freedom and 
describe how the excitation levels of these modes can be selectively controlled 
exploiting the bias dependence of electron-hole pair creation processes \cite{Hartle2010,Volkovich2011b}. 
Such mode-selective vibrational excitation may lead, 
for example, to applications in mode-selective chemistry \cite{Jortner91,Pascual03}, 
where one seeks to break a specific (not necessarily the weakest) chemical bond in a molecule.

To this end, we consider a generic model for a molecular junction with two electronic states and 
two vibrational modes (model MSVE, see Tab.\ \ref{tableII}). 
The two states are located above and below the Fermi level of the junction and may represent, for example, 
the HOMO and the LUMO level of the molecule. Moreover, the two states are coupled more strongly to the 
left than to the right lead. Such a coupling scenario is fairly common, in particular 
if one uses a scanning tunneling microscope to establish the contact to the molecule. 
The two modes are assumed to couple to one of the two states only, 
that is $\lambda_{m,\alpha}\sim\delta_{m\alpha}$. This is an idealized situation, which allows to discuss the 
generic effect. Off-diagonal electronic-vibrational coupling diminishes the effect, which, nevertheless, 
extends over a wide range of parameters (see Ref.\ \cite{Volkovich2011b} for a more detailed discussion, 
including the effect of electron-electron interactions).

\begin{figure} 
\begin{center}
\begin{tabular}{l}
\resizebox{\newwidth}{\newheight}{
\includegraphics{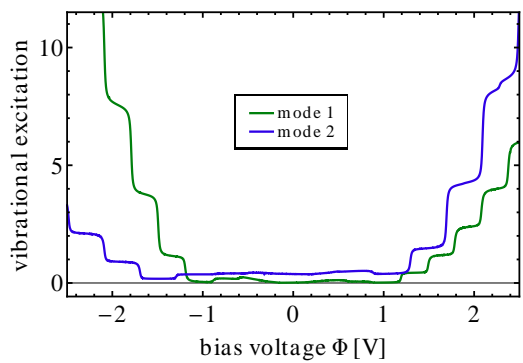}
}
\end{tabular}
\end{center}
  \caption{\label{MSVE-B-el} 
  Vibrational excitation characteristics 
of junction MSVE (cf.\ Tab.\ \ref{tableII}). }
\end{figure}

The vibrational excitation characteristic of the two vibrational modes is depicted in Fig.\ \ref{MSVE-B-el}. 
In particular, mode 1 shows significantly higher levels of excitation than mode 2 for negative and vice versa for positive bias voltages, 
that is, the excitation levels of the two modes can be selectively controlled by the external bias voltage \cite{Hartle2010,Volkovich2011b}. 
This can be explained 
by the same arguments that have been used to analyze the asymmetric excitation levels observed in junction REC (cf.\ Sec.\ \ref{vibrec}), 
because the subsystem consisting of state 1 and mode 1 is effectively decoupled from the subsystem consisting 
of state 2 and mode 2. 
Thereby, it should be noted that the frequency of mode 2 is significantly larger than that of mode 1. 
This means that for positive bias voltages the (current-induced) vibrational energy is distributed in a way that 
is opposite to what statistical arguments would give in equilibrium systems. In more practical terms, 
the stronger bond of the molecule, which is associated with mode 2, may eventually break in this junction 
before the weaker bond. 
Note that such mode-selective vibrational excitation occurs also for 
other molecule-lead coupling scenarios \cite{Volkovich2011b}.

\subsubsection{Temperature dependence of the current in the presence of destructive interference effects}
\label{tempdep}

Electron-hole pair creation processes play also an important role 
if electron transport through a single-molecule junction is influenced by 
quantum interference effects \cite{Hartle2011b,Ballmann2012,Hartle2012}.  
Quantum interference effects in molecular junctions are of interest not only from a fundamental point of view 
but also in the context of technological decive applications, including transistors \cite{Stafford2007}, 
thermoelectric devices \cite{Bergfield2010b} or spin filters \cite{Herrmann2011}. 
Interference effects occur in single-molecule junctions, e.g., when electron transport through the junctions is mediated by  
quasidegenerate electronic states. 
In this section, we consider an example of such a molecular junction that includes a pair 
of quasidegenerate states: a bonding (symmetric) and an antibonding (antisymmetric) state 
(model INT, cf.\ Tab.\ \ref{tableII}). Note that the symmetry of the two states is reflected in the 
different sign of the molecule-lead coupling strength $\nu_{\text{R},2}$.

\begin{figure}
\begin{center}
\begin{tabular}{ll}
(a) & (b) 
\\
\resizebox{\newwidthprime}{\newheightprime}{
\includegraphics{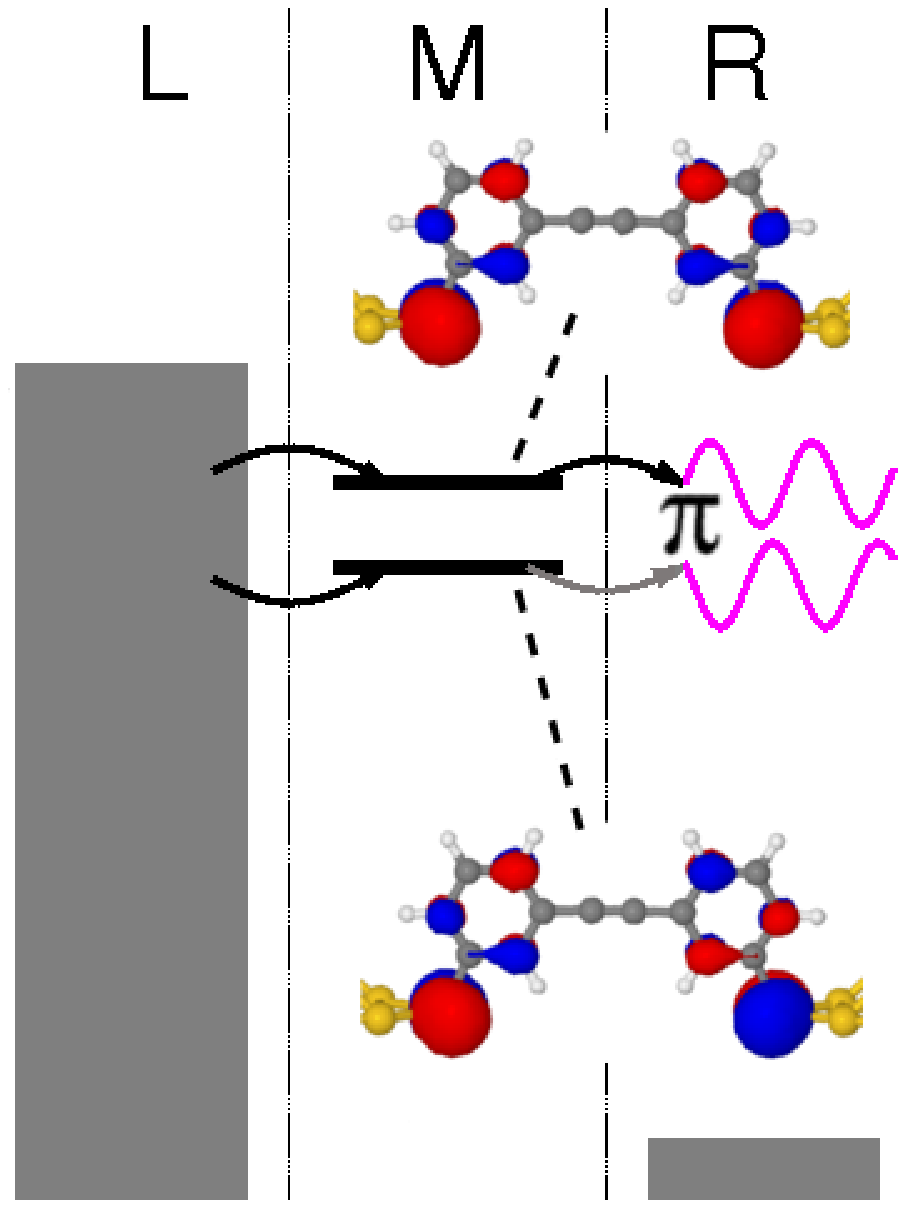}
}
& 
\resizebox{\newwidthprime}{\newheightprime}{
\includegraphics{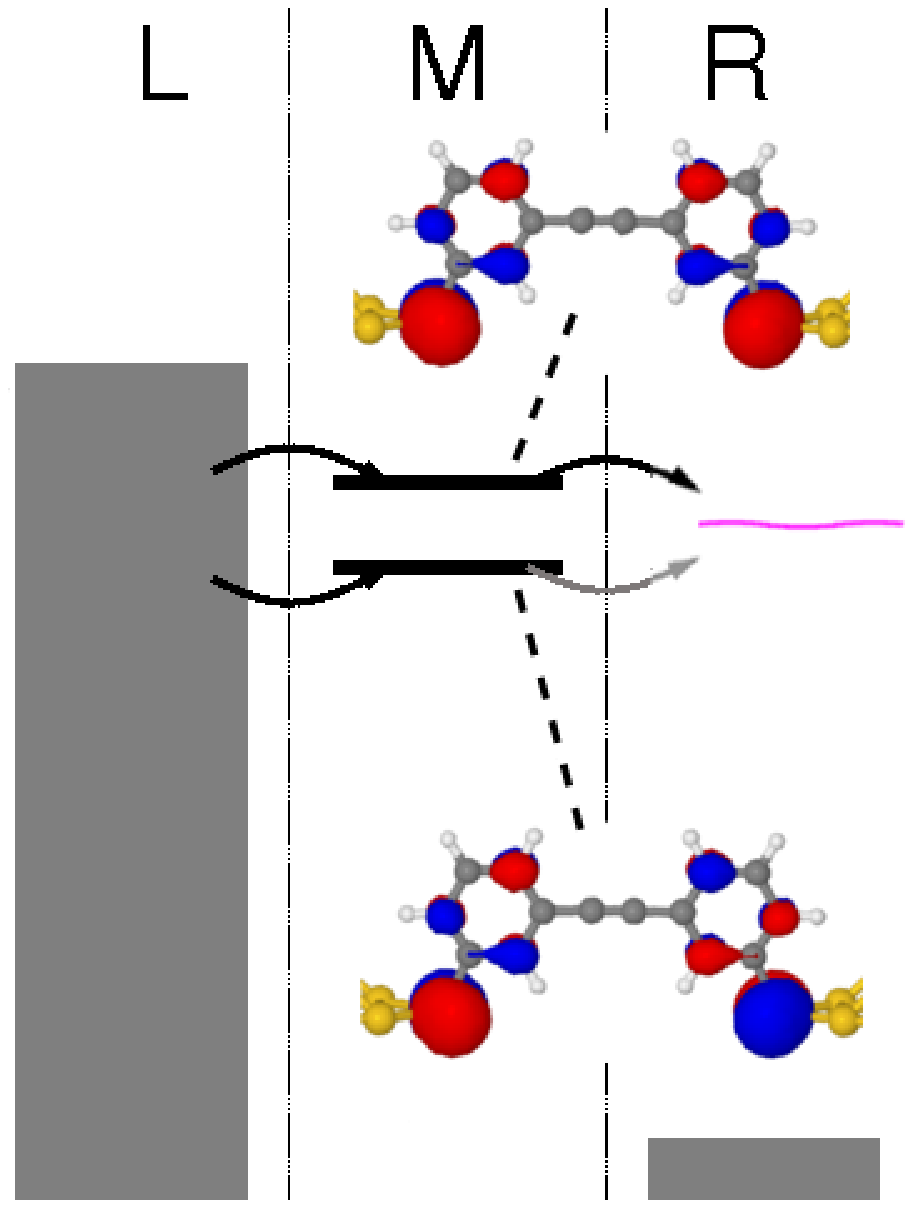} 
}
\end{tabular}
\end{center}
\caption{\label{LinConduct} 
Panel (a) and (b): Graphical representation of a pair of quasidegenerate electronic states. 
Due to the different L$\leftrightarrow$R symmetry of the two states, the tunneling amplitudes for 
(electronic) transport processes through these two states, which 
are depicted by purple wiggly lines, differ by a phase $\pi$. Accordingly, they 
destructively interfer with each other, resulting in a suppression of the corresponding 
tunneling amplitude. 
}
\end{figure}

\begin{figure}
\begin{tabular}{l}
\resizebox{\newwidth}{\newheight}{
\includegraphics{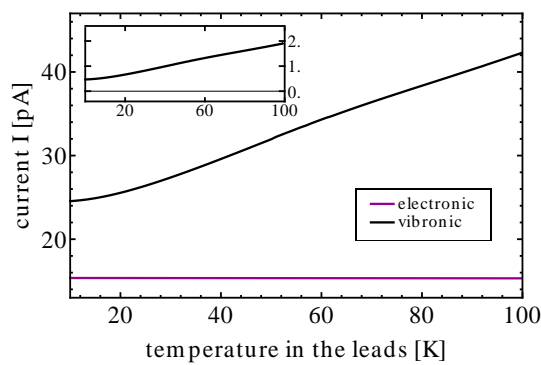}
}
\end{tabular}
\caption{(Color online)\label{FigTempDep} 
Current-temperature characteristics of junction INT (cf.\ Tab.\ \ref{tableII}). 
}
\end{figure}

Due to the different symmetry of the two states, 
the tunneling amplitudes that are associated with electron tunneling through the two states 
differ by a phase $\pi$ and, therefore, destructively interfer 
with each other (see Fig.\ \ref{LinConduct}a and \ref{LinConduct}b). 
This leads to a strong suppression of the tunnel current in this system \cite{Solomon2008b}. 
For example, at $e\Phi=0.2$\,eV, the electronic current level of 
junction INT is given by $\approx15$\,pA 
\footnote{Note that in this section, Sec.\ \ref{tempdep}, we refer to the electronic scenario as to 
the one without electronic-vibrational but including the polaron-shift of the electronic levels. 
Thus, the level of quasi-degeneracy between the electronic states in the vibronic and 
the electronic transport scenario is the same.}, 
while one obtains a current level of $0.4$\,$\mu$A 
if interference effects are discarded in this system.

As we have shown recently, electronic-vibrational coupling can strongly quench  
such interference effects \cite{Hartle2011b,Hartle2012}. Thereby, it is decisive to note that 
the electronic states of a molecular junction couple differently to the vibrational degrees of freedom, 
even if they represent just symmetric and antisymmetric combinations of localized molecular orbitals such 
as, for example, the states depicted in Fig.\ \ref{LinConduct}. Thus, due to the interaction with the vibrational 
degree of freedom, tunneling through one of the states becomes more favorable than through the other 
and, consequently, interference effects are less pronounced. As a result, the current level of junction INT 
is significantly larger than without electronic-vibrational coupling. This can be seen in Fig.\ \ref{FigTempDep}, 
where the solid black and purple line depict the current-temperature characteristic of junction INT 
at $e\Phi=0.2$\,eV for the vibronic and electronic transport scenario, respectively.

Thereby, the current level of the electronic scenario shows no temperature dependence because 
at bias voltages $e\Phi\gg2\overline{\epsilon}_{1}$ thermal broadening in the electrodes has no influence 
on the current level of the junction. In contrast, the vibronic current increases almost linearly with the 
temperature in the electrodes. This temperature dependence is due to the effect of electron-hole pair creation processes 
(Fig.\ \ref{el-h-pair-creation}) \cite{Hartle2012}. As these processes involve only one of the electrodes, they are not suppressed 
by destructive interference effects. Therefore, the level of vibrational excitation is 
determined by electron-hole pair creation processes in this system rather than by inelastic transport processes 
that are suppressed by destructive interference effects. 
Similar as for a molecule that is adsorbed on a surface, pair creation processes adapt the vibrational excitation level of the molecule 
to the temperature in the electrodes, leading to an almost linear increase of vibrational excitation, which 
is shown in the inset of Fig.\ \ref{FigTempDep}, as the temperature 
in the electrodes increases. As a result of the higher excitation levels, 
inelastic transport processes become more favorable, leading to larger current levels in this system due to an enhanced 
quenching of destructive interference effects. 
Thus, electron-hole pair creation processes facilitate a mechanism to control quantum interference effects 
in this junction by an external parameter, \emph{i.e.}\ the temperature in the leads. 
This mechanism has recently been verified in a series of experiments on various single-molecule junctions 
by S.\ Ballmann \emph{et al.}\ \cite{Ballmann2012}.

\section{Conclusions}

The theoretical studies \cite{Hartle,Hartle09,Hartle2010,Hartle2010b,Hartle2011,Hartle2011b,Volkovich2011b,HartlePhD,Hartle2012} 
of vibrationally coupled electron transport reviewed in this paper
show that electron-hole pair creation processes play an important role 
in this nonequilibrium transport problem, although they are not directly contributing 
to the current that is flowing through the junction. Similar to scenarios where a molecule is adsorbed on a surface, 
pair creation processes tend to adapt the vibrational excitation levels of the molecular bridge to the 
temperature in the electrodes. They are directly influencing the vibrational excitation levels of the 
junction and, thus, also the efficiency of transport processes. 
Although this represents an indirect influence on the electrical transport properties of a molecular junction 
it can, nevertheless, lead to a number of interesting transport phenomena such as, for example, 
negative differential resistance \cite{Hartle2010b} and rectification \cite{Hartle2010b,Volkovich2011b}. 
In contrast to transport processes, electron-hole pair creation processes 
involve only one of the electrodes. As a consequence they transfer an asymmetry in the molecule-lead coupling 
to the corresponding transport characteristics. This is of relevance for spectroscopic 
applications of single-molecule junctions \cite{HartlePhD}, 
mode-selective vibrational excitation \cite{Hartle2010,Volkovich2011b} 
or in the presence of destructive quantum interference effects \cite{Ballmann2012,Hartle2012}. 
Moreover, it was shown that the suppression of electron-hole pair creation processes leads to higher 
levels of vibrational excitation in systems with weaker electronic-vibrational coupling \cite{Hartle2010b} and that the 
absence of electron-hole pair creation processes represents a link between the limit of high bias voltages 
and weak electronic-vibrational coupling (for $\Phi>2(\overline{\epsilon}_{1}+\Omega_{1})$) \cite{Hartle2011}.

Throughout the article, we have focused on resonant processes. It should be noted though, that, 
similar to resonant and non-resonant transport processes \cite{Lueffe,Hartle}, 
off-resonant electron-hole pair creation processes are also of importance
in molecular junctions. Aspects and implications of these processes are discussed, for example, 
in Ref.\ \cite{Volkovich2011b}, where it is shown that they play an important role 
in the non-resonant transport regime of a molecular junction but also at high bias voltages.

\section*{Acknowledgement}

We thank R.\ Volkovich, S.\ Wagner, S.\ Ballmann, 
O.\ Godsi, D.\ Brisker Klaiman, S.\ Klaiman,  
M.\ Butzin, I.\ Pshenichnyuk, O.\ Rubio-Pons, P.\ B.\ Coto, B.\ Kubala, M.\ Cizek, A.\ Nitzan and 
H.\ B.\ Weber for many fruitful and inspiring discussions. 
The generous allocation of computing time by the computing centers 
in Erlangen (RRZE), Munich (LRZ), and J\"ulich (JSC) is gratefully acknowledged. 
This work has been supported by the 
German-Israeli Foundation for Scientific Development (GIF) and the 
Deutsche Forschungsgemeinschaft (DFG) 
through SPP 1243, the Cluster of Excellence 'Engineering of Advanced Materials' and SFB 953. 
MT gratefully acknowledges the hospitality of the Institute of Advanced Studies at the Hebrew University Jerusalem 
within the workshop on molecular electronics and the Pitzer Center for Theoretical Chemistry 
of the University of California at Berkeley. 
RH is grateful for financial support by the National Science Foundation (DMR-1006282) 
and the Alexander von Humboldt Foundation.

\bibliographystyle{pss}
\providecommand{\WileyBibTextsc}{}
\let\textsc\WileyBibTextsc
\providecommand{\othercit}{}
\providecommand{\jr}[1]{#1}
\providecommand{\etal}{~et~al.}

\end{document}